\documentclass[10pt]{elsart_MofM}
\makeatletter
  \renewcommand{\@overcaptionskip}{5mm}

  \def\appendix@nostar{%
    \def\lb@section{\appendixname\ \thesection.\half@em}
    \def\lb@empty@section{\appendixname\ \thesection}
    \setcounter{section}{0}\def\thesection{\Alph{section}}%
    \setcounter{subsection}{0}%
    \setcounter{subsubsection}{0}%
    \setcounter{paragraph}{0}%
    \app@number{equation}}
  \def\description{\list{}{%
    \leftmargin 12mm \topsep 2ex
    \listparindent 2em
    \let\makelabel\descriptionlabel}}

  \def\enumerate{%
    \ifnum \@enumdepth >\@maxlistdepth
      \@toodeep
    \else
      \advance\@enumdepth \@ne
      \edef\@enumctr{enum\romannumeral\the\@enumdepth}%
      \list{\csname label\@enumctr\endcsname}%
         {\usecounter{\@enumctr}
         \leftmargin 12mm \topsep 1ex
         \listparindent 2em
         \let\makelabel=\right@label}
    \fi}
  \def\itemize{%
    \ifnum \@itemdepth >\@maxlistdepth
      \@toodeep
    \else
      \advance\@itemdepth \@ne
      \edef\@itemitem{labelitem\romannumeral\the\@itemdepth}%
       \setleftmargin{i}{--}%
       \setleftmargin{ii}{$\cdot$}%
      \list{\csname\@itemitem\endcsname}%
         {%
          \leftmargin 12mm \topsep 1ex
          \listparindent 2em
          \let\makelabel\right@label}
    \fi}
  \long\def\@maketablecaption#1#2{\@tablecaptionsize
    \global \@minipagefalse
    \hbox to \hsize{\parbox[t]{\hsize}{#1 \\ #2 \\[-1ex]}}}
  \def\section{\@startsection{section}{1}{\z@}{1.0\@bls
    \@plus .4\@bls \@minus .1\@bls}{0.5\@bls}{\normalsize\bfseries}}
  \def\subsection{\@startsection{subsection}{2}{\z@}{1.0\@bls
    \@plus 0.\@bls \@minus .1\@bls}{0.4\@bls}{\normalsize\itshape}}
  \def\subsubsection{\@startsection{subsubsection}{3}{\z@}{0.5\@bls
    \@plus 0.\@bls}{0.0001pt}{\normalsize\itshape}}
  \def\@titlesize{\large\bfseries}
  \def\@authorsize{\normalsize}
\makeatother
\usepackage{graphicx}
\usepackage[footnotesize,sf]{subfigure}

\DeclareMathAlphabet{\mathpzc}{OT1}{pzc}{m}{it}
\newcommand{\PC}{\mathpzc{pc}}
\newcommand{\QC}{\mathpzc{qc}}
\usepackage{chicago}
\usepackage{dcolumn}
\newcolumntype{d}{D{.}{.}{-1}}
\usepackage{amsmath}
\usepackage{amssymb}
\begin{document}
\begin{frontmatter}
%
\title{Contact Rolling and Deformation in Granular Media}
\author[USA]{Matthew R. Kuhn\corauthref{cor}},
  \ead{kuhn@up.edu}
\author[Hungary]{Katalin Bagi}
  \ead{kbagi@mail.bme.hu}
\address[USA]{Dept.\ of Civil and Env.\ Engrg.,
  School of Engrg., Univ.\ of Portland, \\
  5000 N.\ Willamette Blvd., Portland, OR\ 97203, U.S.A. \\
  Fax. 1-503-943-7316}
\address[Hungary]{Research Group for Computational
                  Structural Mechanics, Hungarian Academy of Sciences,\\
                  Dept.\ of Structural Mechanics,
  Technical University of Budapest,\\
  M\H{u}egyetem rkp.~3, K.mf.~35, H-1521 Budapest, Hungary}
\corauth[cor]{Corresponding author.}
%
%
\begin{abstract}
\small
The paper considers rotations at different scales in granular materials:
the rotations of individual particles, the rolling and rigid-rotation
of particle pairs, 
the rotational interactions of a particle within its cluster of neighbors,
and the rotation of material regions.
Numerical, Discrete Element Method (DEM) simulations on
two- and three-dimensional (2D and 3D) assemblies show that
particle rotations are diverse, that they increase with strain until
the material begins to soften, and that they are expressed in
spatial patterns, even at small strains.
The interactions of a pair of particles are a combination of
three modes: a contact deformation mode, a contact rolling mode,
and a mode of rigid pair motions.
Definitions are presented for each mode, including four
different
definitions of contact rolling.
A rolling curl is also defined, which describes
the cumulative rolling of neighboring particles around a central
particle or sub-region.
At a larger scale, material deformation and rotation 
are measured within small sub-regions of material, and the
material deformation can be attributed to separate
contributions of contact rolling, contact deformation, and the rigid-rotation
of particle pairs.
The diversity and extend of contact rolling
were measured in 2D and 3D simulations.  
A dominant rolling pattern was observed, which resembles the interactions
of rolling gears.
This pattern can extend to distances of at least six particle diameters
from a central particle.
\end{abstract}
\end{frontmatter}
\small
%
\section{Introduction} \label{sec:introduction}
Particle rotations are known to have a fundamental influence on the behavior of
granular materials.  In presenting their results of experiments on
plastic rods, \shortciteN{Oda:1982a} 
concluded that ``particle rolling is indeed a
dominant microscopic feature, especially in the presence of inter-particle
friction.''  Subsequent experiments have sought to
characterize the manner and extent of particle rotations and their effects upon
the mechanical behavior of granular materials,  
and much of this work is reviewed.
The current study 
characterizes particle rotations and rolling
in granular materials by considering a hierarchy of rotational effects:
the rotations of individual particles, the rotational interactions
of particle pairs, the rotational interactions of a particle within its
cluster of neighbors, and the rotations of larger material regions.
Although our intent is not to develop a comprehensive account of particle
rotations and their effects, we will attempt to clarify certain aspects of
behavior and to fill in gaps in the current understanding.  
Our work is primarily experimental, 
and it is based upon rational definitions of such notions as 
granular deformation, rolling, material rotation, and material curl
when applied to a discrete, granular media.
Discrete Element Method (DEM) simulations are presented
as a means of exploring and quantifying granular deformation
and rolling in a realistic setting.
The simulations are both two- and three-dimensional (2D and 3D), and they
were conducted on circular, non-circular, spherical, and
non-spherical particle assemblies.
\par
The paper is organized in the following manner, with
our original developments appearing primarily in 
Sections~\ref{sec:pairs} to~\ref{sec:def-rot}:
\begin{itemize}
\item
In Section~2, we consider the rotations of particle units.  We 
describe the experiments that are cited throughout the
paper, and we corroborate the results of previous studies concerning
statistical measures of particle rotations.
We also illustrate a form of spatial patterning in the particle rotations.
\item
In Section~3, we consider the interactions of particle pairs.
We summarize and apply previous work by the authors, in which we have
distinguished and defined three forms of contact interactions: 
contact deformation, contact rolling (which we define in alternative ways),
and rigid-rotation of the particle pair.
All three motions are usually active at each contact at
the same time, and we present
experimental results that quantify the various interactions
and their inter-relationships.
We consider the manner in which particle rotations and rolling affect
material behavior, and we illustrate a dominant pattern in the
rolling among particles.
\item
In Section~4, we define a micro-measure of the rolling between a particle and 
its cluster of nearest neighbors.
This measure, akin to the curl in a continuous media, is measured 
in experiments, and the results are compared with those in Section~3.
\item
In Section~5, we derive expressions for the deformation and rotation of discrete
material sub-regions of 2D assemblies.
The material rotations are compared with the particle rotations,
and the material deformation is partitioned so that we
can measure the separate effects of contact
deformation, contact rolling, and the rigid-rotation of particle pairs.
\end{itemize}
\section{Particle rotations}\label{sec:particles}
As a granular material is deformed, grains interact with each
other through their contacts.  These interactions
are produced by the translations and rotations of the grains.
In this section, we consider the rotations of the individual
grains, statistical measures of these rotations, and the patterning
of these particle rotations.
In later sections, we discuss the rotational interactions of
particle pairs, clusters, and regions.
Particle rotations have been measured in 2D laboratory models,
in numerical (DEM) simulations, and in 3D physical experiments
(for example, \shortciteNP{Oda:1982a,Bardet:1994a,%
Calvetti:1997a,Daudon:1997a,Misra:1997a,Lanier:2001a,Marcher:2001a,%
Matsushima:2003a}).
Experiments have consistently shown that, although particle rotations may
be very large, the mean rotation of the particles within
a large assembly is nearly equal to the mean, continuum spin of the
assembly.
The equivalence of mean-field
rotation and the particle rotation has also been reported under conditions of
non-uniform shearing.  
For example, \citeN{Bardet:1994a} conducted biaxial DEM
simulations on a 2D assembly that had flexible side boundaries and rigid upper
and lower platens.  
A persistent shear band spontaneously appeared under these
conditions.  Although particle rotations within the shear band were large and
had a predominant direction, the mean particle rotation of the entire assembly
was nearly zero.  
In their DEM tests,
\shortciteN{Matsushima:2003a} found that the mean
grain rotation and the continuum rotation
were nearly equal, even within a shear band. 
\citeN{Kuhn:1999a} performed DEM shearing tests on a 2D
assembly of circular disks and produced highly non-uniform patterns of shearing
by applying body forces on the assembly.  
Even under conditions of large
gradients in the shearing strain, the particle rotations were, on average,
nearly equal to the mean-field rotation.  
Two studies have shown a difference, however, between the mean particle and
mean-field rotations.
In the special 2D simulations of
\shortciteN{Calvetti:1997a} 
the mean particle rotation was found to drift from the
mean-field rotation.  
During those tests, the principal stress axes were
continually rotated, which produced a progressively larger drift in the average
particle rotation.
Recently, \citeN{Jenkins:2003a}
reported that in their triaxial compression simulations,
the average spin of particles differed from the overall assembly rotation.
\par
Although the conformance of the mean-field and mean-particle rotations may
suggest a certain order among the rotations, 
this order is contradicted by the
consistent observation of substantial variability among individual rotations.
Experiments have shown that particle rotations are large 
and that the magnitude
and variation of rotations increases with 
increasing strain \shortcite{Oda:1982a,%
Bagi:1993a,Bardet:1994a,Misra:1997a,Calvetti:1997a,Lin:1997a,Dedecker:2000a,%
Lanier:2001a,Kuhn:2002a}.  
Even particle rotations in regular, hexagonal
packings of coins and plastic rods have been found to be both large and diverse
\shortcite{Tamura:1998a,Khidas:2001a}.  
Investigators have used the statistical
standard deviation of the particle rotations as a measure of their magnitude
and of the fluctuation of individual rotations from the mean.  
As an example,
\shortciteN{Dedecker:2000a} 
measured the particle spin rates in numerical biaxial
tests on circular particles.  
At large compressive strains, the standard
deviation of the particle spin was as large as twenty times the strain rate.
In experiments on wood rods, \shortciteN{Calvetti:1997a} also observed large
fluctuations from the mean and found that the standard deviation of particle
rotations increased in a consistent and nearly linear manner with increasing
strain.  
Although evidence of rotation variability is now abundant at larger
strains, the experimental record provides less evidence of the variability of
particle rotations at small strains.  
This scarcity is likely due to the
difficulty of measuring the initial particle velocities in physical experiments.
\subsection{Computer simulations}\label{sec:simulations}
To investigate the mechanical roles of rotation, rolling, and deformation,
we conducted numerical, DEM simulations on five 
large two- and three-dimensional
assemblies of densely compacted particles (see \citeNP{Cundall:1979a}
for algorithm details).
Although our primary focus will be on the more realistic 3D simulations,
we conducted experiments on simpler 2D assemblies 
so that spatial patterning could be more easily observed, and so
that we could directly measure the function of inter-particle rolling in
the deformation of granular media.
\par
Two 2D assemblies were tested: an assembly of disks and an assembly of
ovals, each with 10,816 particles (Table~\ref{table:assemblies}).
Three 3D assemblies were also tested, each containing 4096 spheres or
non-spherical (oblate or prolate) solids of revolution called \emph{ovoids}.
The oval and ovoid shapes are smooth and convex, closely resembling
ellipses and spheroids,
and their construction and numerical treatment
are described elsewhere~\cite{Potapov:1998a,Kuhn:2003a}.
These shapes had the
following aspect ratios:  ovals, 1.35; oblate ovoids, 0.65--0.85;
and prolate ovoids, 1.2--1.6.
In all assemblies, a range of particle sizes was used: 
0.5$\overline{D}$--1.7$\overline{D}$
for the 2D assemblies, and 0.5$\overline{D}$--1.35$\overline{D}$
for the 3D assemblies, where $\overline{D}$ is the mean particle size.
\par
The dense square and cubic assemblies were compacted from initially
sparse arrangements of particles.
The sparse assemblies were isotropically densified by 
converging their periodic boundaries.
Densification was also promoted by ``turning off'' contact friction
during this process (as in \citeNP{Thornton:2000b,Roux:2003a}) and
by periodically energizing 
the assembly by assigning random particle velocities.
Table~\ref{table:assemblies} shows
the initial void ratio, solid fraction, and average coordination
number of each assembly.
\begin{table}
\centering
\caption{Initial states of five assemblies.}
\begin{tabular}{lccccccc}
\hline
  &\multicolumn{2}{c}{2D assemblies}&& \multicolumn{3}{c}{3D assemblies} \\
\cline{2-3}\cline{5-7}
                      &  Circles & Ovals && Spheres & Oblate & Prolate \\
\hline
Number of particles   &  10816   & 10816 && 4096    & 4096   & 4096    \\
Void ratio, $e$       &  0.173   & 0.112 && 0.509   & 0.376  & 0.372   \\
Solid fraction        &  0.853   & 0.899 && 0.663   & 0.727  & 0.729   \\
Avg. coordination no. &  8.82    & 5.59  && 5.57    & 8.97   & 9.24    \\
\hline
\end{tabular}
\label{table:assemblies}
\end{table}
\par
All assemblies were tested in either biaxial (2D) or triaxial (3D)
compression within their periodic boundaries:
the assembly height was slowly reduced while keeping a constant
average normal stress along the sides. 
During the loading tests a simple force mechanism
was employed between contacting particles.
Linear, elastic normal and tangential contact springs were assigned equal
stiffnesses ($k_{\mathrm{n}}=k_{\mathrm{t}}$),
and slipping between particles would occur whenever
the contact friction coefficient of 0.50 was attained.
Although the contact characteristics between solid
granules are likely closer to Hertzian than to linear, which may affect
the rotational behaviour \cite{Jenkins:2003a}, 
our present study explores
the relative intensities of the rolling motions without any pretense
of replicating the behavior of a particular material.
Rolling and rotation were allowed to occur freely at the contacts.
\par
\begin{figure}
\centering
\mbox{%
\subfigure[Two-dimensional assemblies]%
  {\includegraphics[scale=0.80]{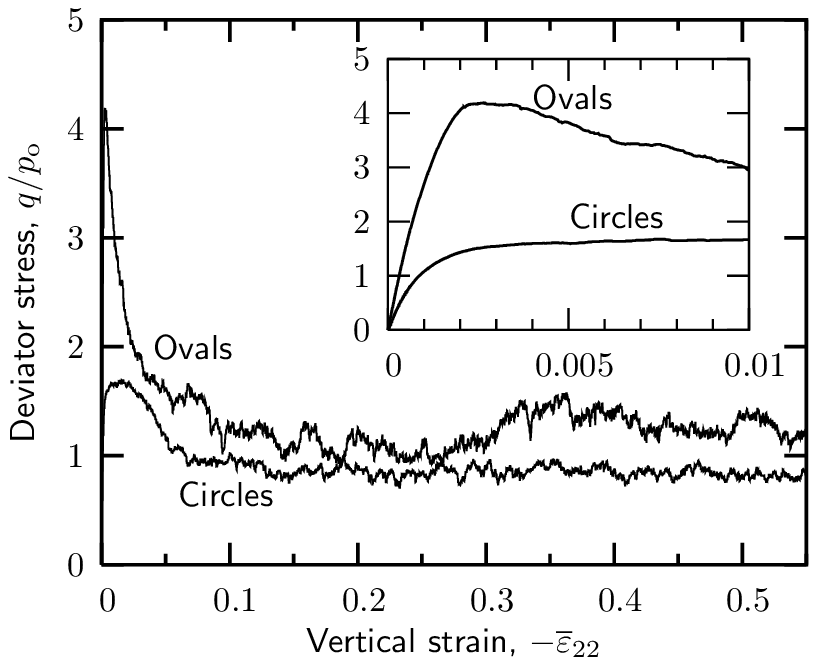}}%
\quad%
\subfigure[Three-dimensional assemblies]
  {\includegraphics[scale=0.80]{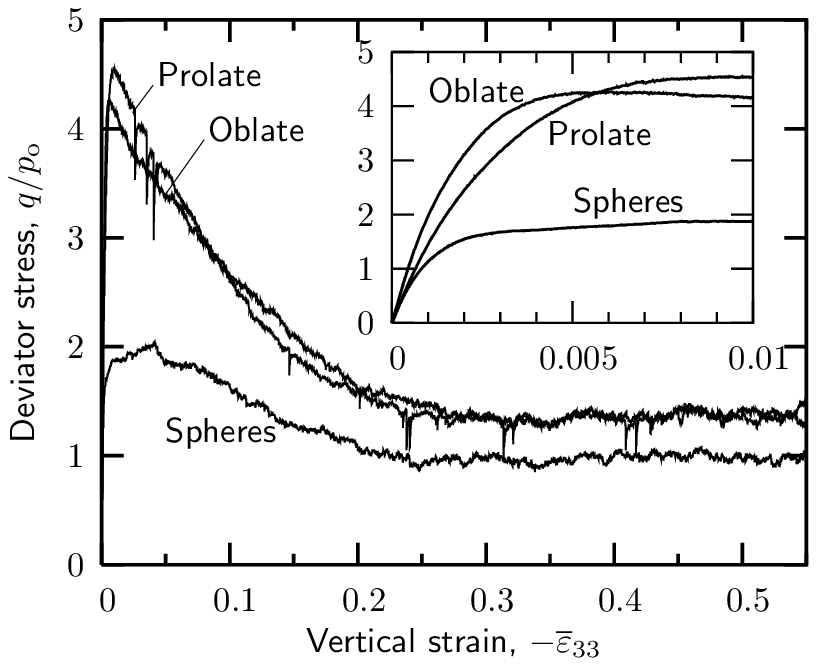}}%
}
\caption{Loading behavior of all assemblies in biaxial or triaxial compression.}
\label{fig:crs}
\end{figure}

Figure~\ref{fig:crs} shows the evolution of the deviator stress
$q=\sigma_{33} - \sigma_{11}$ (or $\sigma_{22} - \sigma_{11}$ in 2D)
during loading,
where $q$ is expressed in a dimensionless form by
dividing by the initial (negative) mean stress~$p_{\text{o}}$.
The strains $\epsilon_{22}$ and $\epsilon_{33}$
in the figure
are Lagrangian, engineering strains, computed
as the change in the assembly's height divided by its
initial height.
Elsewhere in the paper we employ an
Eulerian strain increment, $d\epsilon_{22}$ or $d\epsilon_{33}$,
as a small change in height relative to the current height.
In order to analyze the incremental interactions of the particles,
the positions of all particles
were periodically stored at two
nearby states, separated by a small strain
increment $d\epsilon_{33}=-5\times 10^{-5}$. 
Pairs of these states were later used to
compute the incremental rates of particle rotation, contact rolling,
and other quantities.
\par
The observed behavior in compression is typical for densely packed, unbonded
granular materials (Fig.~\ref{fig:crs}).
The response is initially elastic, but plastic deformation soon
dominates as the deviator stress approaches its peak state.
The material is strongly dilatant at the peak state and during subsequent
softening.
At a vertical strain $-\epsilon_{33}$ of about 0.30,
the material reaches a steady (critical) state of nearly constant
stress and volume.
Localization is present in the 2D assemblies in
the form of micro-bands at small strains
and as shear bands at the peak state and beyond.
Although we did not search for localization in the 3D assemblies,
shear bands are unlikely, as the assembly was a fairly small cube, 
with a width less than 16 particle diameters.
\subsection{Simulation results: particle rotations}\label{sec:part-rot}
For conditions of biaxial and triaxial compression,
the mean (continuum) rotation is zero, and our experiments
show that the mean rotation of the particles, although not zero,
is very small.
This result is evident in the second column (``Averaged rotation'')
of Table~\ref{table:part-rot}, which gives the averages
of the mean particle rotations for the five assemblies at three strains.
\begin{table}
\centering
\caption{Mean and standard deviations of the particle rotation rates
$d\boldsymbol{\theta}^{p}$
for five assemblies.}
\begin{tabular}{lcccc}
\hline
               & Averaged & \multicolumn{3}{c}
{$\mathsf{Std}(d\theta^{p}_{i})/|d\epsilon_{33}|\:^{\text{a}}$}\\
\cline{3-5}
Assembly       & rotation & Zero & Peak & Steady \\
\hline
Circles        & 0.10     & 0.96 & 19.5 & 32.0 \\
Ovals          & 0.11     & 0.24 & 26.6 & 24.5 \\
Spheres        & 0.10     & 0.65 & 12.4 & 16.3 \\
Oblate ovoids  & 0.07     & 0.28 & 12.3 & 11.0 \\
Prolate ovoids & 0.10     & 0.26 & 9.1  & 13.2 \\
\hline
\multicolumn{5}{l}{$^{\text{a}}$\ \parbox[t]{9cm}{%
For two-dimensional assemblies,
rotations $d\theta^{p}_{3}$ are reported; for three-dimensional assemblies,
rotations $d\theta^{p}_{1}$ are reported.}}\\
\hline
\end{tabular}
\label{table:part-rot}
\end{table}
Each average is over the entire range of strains of a simulation
and for the magnitude of the mean particle rotation
at each strain:
\begin{equation}\label{eq:meanmean}
\text{Averaged rotation} \equiv
\underset{\epsilon_{33} = 0\text{ to }0.50}{\mathsf{Mean}}
\left(
\left|
\underset{p = 1\text{ to }N}{\mathsf{Mean}}
(d\boldsymbol{\theta}^{p} / |d\epsilon_{33}|)
\right|
\right)
\;,
\end{equation}
where $p$ is the particle index, $d\boldsymbol{\theta}^p$ is
a particle rotation, and $N$ is the number of participating particles
in the assembly (either 4096 or 10,816, Table~\ref{table:assemblies}).
Because the simulations did not include the
influence of gravity, many particles are without contacts,
and these non-participating, ``floating'' particles 
are excluded in our analysis of incremental effects.
The particle rotations are expressed in a normalized form by
dividing by the strain increment $d\epsilon_{33}$ over which they are measured.
Although the mean particle rotations are small, they are not zero.
The small deviation is likely the result of averaging 
a population of individual 
rotations which exhibit an extreme variability.
The mean rotation 
is certainly small when measured against the variability
of the rotations:
the mean rotation is 
typically less than one hundredth of the standard
deviation of the rotations, as suggested by the 
standard deviations in the final two columns
of Table~\ref{table:part-rot} 
(the average in Eq.~\ref{eq:meanmean} 
is dominated by the range of strains between
the peak and steady states).
\par
\begin{figure}
  \centering
  \includegraphics[scale=0.9]{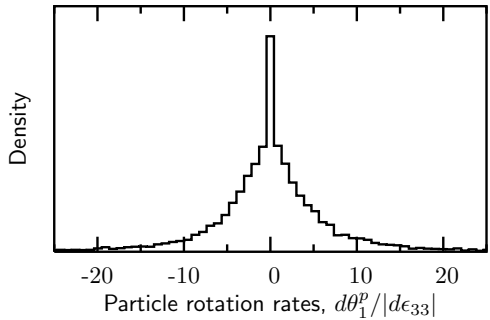}%
  \caption{Probability density of the particle rotation rates
           $d\theta^{p}_{1}/|d\epsilon_{33}|$ in the assembly of
           oblate ovoids at the peak stress condition.}
  \label{fig:hist-rot}
\end{figure}

The particle rotations exhibit large fluctuations from their mean.
Figure~\ref{fig:hist-rot} shows the distribution of particle rotation rates
$d\theta_{1}$ within
the oblate ovoid assembly at the peak stress.
The standard deviation of the normalized rate $d\theta_{1}/|d\epsilon_{33}|$
is about 12, and over 5\% of the particles are rotating 
20 times faster than the vertical strain rate. 
Standard deviations of the rotation rates are summarized in
the final three columns of Table~\ref{table:part-rot} for the five assemblies
at zero strain, at the peak stress, and during steady state deformation.
Rotation rates at the peak state are from 20 to over 100 times
greater than those at zero strain, but rotation rates are similar at the
peak and steady states. 
These large increases in the particle rotation rates
are consistent with the results of \shortciteN{Calvetti:1997a},
but we find that the rotation rates do not increase substantially 
after the peak strength has been reached.
As would be expected,
rotations are smaller at zero strain
with non-spherical shapes than with spheres.
At large strains, however, the rotation rates are similar for all 
three shapes.
\subsection{Simulation results: spatial patterning of particle rotations}
\label{sec:strings}
\begin{figure}
\centering
\mbox{%
\subfigure[Circles, $\epsilon_{22}=0$]%
  {\includegraphics[scale=0.90]{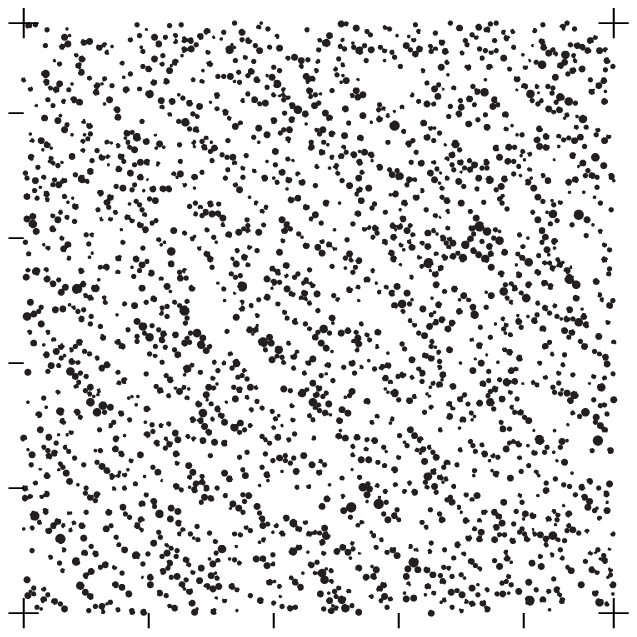}}%
\quad%
\subfigure[Ovals, $\epsilon_{22}=\mbox{}-0.0012$]
  {\includegraphics[scale=0.90]{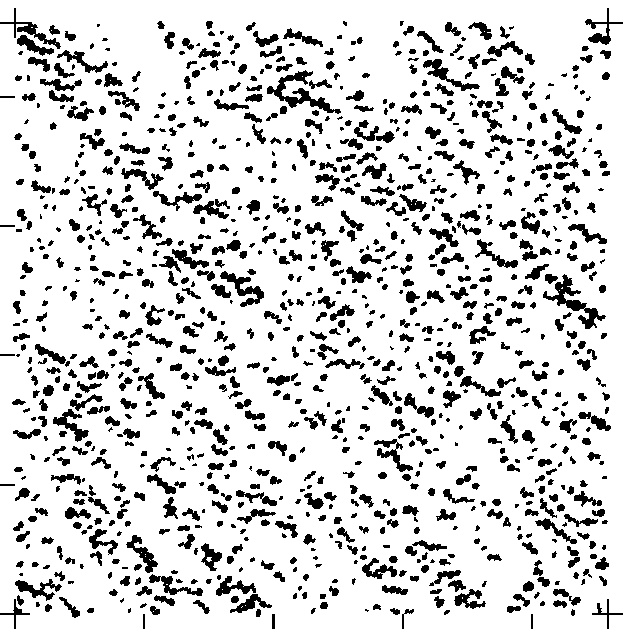}}%
}
\caption{Rotation rates of circles and ovals at two strains during
biaxial compression.
The figures only show particles that are rotating counter-clockwise.
In (a), the shaded circles have rotation rates that exceed
$d\theta_{3}/d\epsilon_{22} = 0.6$; 
in (b), the shaded ovals have rotation rates that
exceed $d\theta_{3}/d\epsilon_{22} = 1.0$.}
\label{fig:strings}
\end{figure}

Figure~\ref{fig:strings} shows the spatial
distribution of the counter-clockwise rotations of
particles in the assemblies of circles (Fig.~\ref{fig:strings}a)
and ovals (Fig.~\ref{fig:strings}b).
Only counter-clockwise rotations are shown in these monochrome plots,
where the shading depends upon the dimensionless rotation
rate $d\theta_{3}/|d\epsilon_{22}|$.
Figure~\ref{fig:strings}a gives disk rotation rates at the start of biaxial
compression ($\epsilon_{22}=0$);
Fig.~\ref{fig:strings}b shows oval rotation rates at the hardening
strain $\epsilon_{22}=-0.0012$, when the deviator
stress had reached about 70\% of its peak value.
Both figures show a pervasive feature of granular rotation at
small strains:
the most rapidly rotating particles are usually
aligned in chain-like patterns oblique to the principal stress directions.
These \emph{rotation chains} are somewhat more sinuous at the larger strain,
but allowing for their frequent crooks and staggers, 
some rotation chains can be
traced across the full height and width of the assembly and join
other chains across the periodic boundaries.  
The particles within
counter-clockwise rotation chains are usually not touching each other, as this
would produce intense sliding between particles.
The chains are closely associated with microbands---thin 
granular regions of more intense shearing strain and
dilation \cite{Kuhn:1999a}.
These chains are also observed in the rolling curl
plots of Section~\ref{sec:sim-pat-curl}.
After the appearance of a shear band, the rotation chains are obscured
by the more intense rotations within the shear band, 
although rotation chains can also trend through a shear band.
\section{Interactions of particle pairs}\label{sec:pairs}
The motions of a contacting pair
of particles can be characterized as a combination
of three modal forms: a deformation mode that produces sliding
and indentation at the contact, a rolling interaction, and
a rigid body rotation (rigid-rotation) and translation of the pair.
In this section, we define each of these three modes and we present
experimental results that characterize their nature and relative importance.
\subsection{Definitions of contact deformation, rolling, and rigid motion}%
\begin{figure}
  \centering
  \includegraphics[scale=0.90]{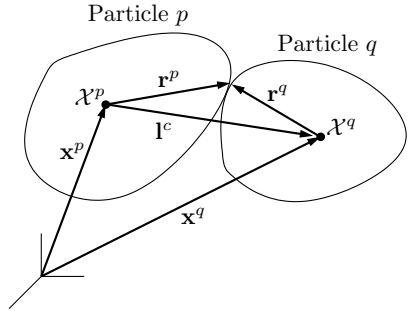}%
  \caption{Two contacting particles.}
  \label{fig:particles}
\end{figure}

\label{sec:definitions}
Consider two particles, $p$ and $q$, to which we assign the material
reference points $\boldsymbol{\chi}^{p}$ and $\boldsymbol{\chi}^{q}$
(Fig.~\ref{fig:particles}).
These reference points are located at positions $\mathbf{x}^{p}$
and $\mathbf{x}^{q}$ relative to the global axes.
The two particles are in contact,
and the vectors $\mathbf{r}^{p}$ and $\mathbf{r}^{q}$ connect 
points $\boldsymbol{\chi}^{p}$ and $\boldsymbol{\chi}^{q}$
to the contact. 
Branch vector $\mathbf{l}$ connects
the two reference points:  
$\mathbf{l} = \mathbf{r}^{p}-\mathbf{r}^{q} = \mathbf{x}^{q}-\mathbf{x}^{p}$.
The contact is assumed to be point-like, with a contact area that is
negligible compared to particle size.
The particles undergo incremental translational and rotational
movements $d\mathbf{u}^{p}$, $d\mathbf{u}^{p}$,
$d\boldsymbol{\theta}^{p}$, and $d\boldsymbol{\theta}^{q}$
during time increment~$dt$.
These four movement vectors are described by twelve scalar components,
which form a 12-dimensional vector space of possible movements.
In classical kinematics, the particles might be truly rigid,
and, hence, prevented from inter-penetration (a non-holonomic constraint),
but the behavior of granular materials is influenced by the
local particle deformations at their contacts.
The contact deformations 
$d\mathbf{u}^{\text{def}}$
are produced by the relative displacements
of two material points, one on either side of the contact,
\begin{equation} \label{eq:udef}
d\mathbf{u}^{\text{def}} =
\left(
d\mathbf{u}^{q} - d\mathbf{u}^{p}
\right)
 +
\left(
d\boldsymbol{\theta}^{q} \times \mathbf{r}^{q} -
d\boldsymbol{\theta}^{p} \times \mathbf{r}^{p}
\right)\;.
\end{equation}
We refer to this displacement as a \emph{contact deformation}, and it
can be separated into components
that are normal and tangent to the contact surface,
producing indentation and sliding.
The contact deformation could also be termed a ``contact displacement,''
but it is distinct from the deformation of a material region,
which is produced by the contact deformations and also by the
rearrangements of the particles (Section~\ref{sec:def-rot}).
The contact deformation~(\ref{eq:udef})
has long been used in the analysis and simulation of granular media
\cite{Cundall:1979a,Molenkamp:1984a,Koenders:1987a,Chang:1989b},
but the sub-space of
deformation motions accounts for only three of
the twelve degrees of freedom of a particle pair.
The deformation motion is objective, since its scalar components
would be properly reported by two observers having
independent motions~\cite{Kuhn:2004b}.
Because the contact deformation is objective, it may be used
in a constitutive description of the contact force--deformation relationship.
\par
Although contact deformation can be unambiguously defined, contact
rolling can be construed in alternative ways.
Several notions of rolling have already been advanced
(as in \shortciteNP{Oda:1982a,Molenkamp:1984a,Bardet:1994a,Iwashita:1998a,%
Lanier:2001a}),
but these definitions are not generally applicable to 2D or 3D
particles of arbitrary shape.
We start with a general view of rolling, which admits
numerous alternative forms:  we view a contact
rolling motion as any objective motion of a particle pair that
is distinct from the contact deformation in Eq.~(\ref{eq:udef}).
We have proposed three definitions of contact rolling,
which are reviewed in the following paragraphs
\cite{Bagi:2004a,Kuhn:2004b}.
A fourth definition is also described.
\par
The simplest definition of contact rolling is based on
the relative rotation of a particle pair,
\begin{equation} \label{eq:relrotation}
d\boldsymbol{\theta}^{\text{rel}} =
d\boldsymbol{\theta}^{q} - d\boldsymbol{\theta}^{p}\;,
\end{equation}
and, like the contact deformation motion in Eq.~(\ref{eq:udef}),
the relative rotation~(\ref{eq:relrotation})
is clearly objective.
The motion $d\boldsymbol{\theta}^{\text{rel}}$ can be separated into
two components, one aligned with the contact
normal $\mathbf{n}$ and the other perpendicular to
$\mathbf{n}$:
\begin{align}
\label{eq:twist}
d\boldsymbol{\theta}^{\text{rel, twist}} &=
(d\boldsymbol{\theta}^{\text{rel}}\cdot\mathbf{n})\mathbf{n}\\
\label{eq:roll1}
d\boldsymbol{\theta}^{\text{rel, roll, 1}} &=
d\boldsymbol{\theta}^{\text{rel}} -
(d\boldsymbol{\theta}^{\text{rel}}\cdot\mathbf{n})\mathbf{n}\; ,
\end{align}
where the contact normal $\mathbf{n}$ is directed outward from
particle $p$.
The index ``1'' denotes the name \emph{Type~1 rolling}, as described by
the authors~\cite{Kuhn:2004c}.
The four motions~(\ref{eq:udef})--(\ref{eq:roll1}) would be
assigned opposite (negative) values if the indices $p$ and $q$
are exchanged.
When plotting graphic visualizations of simulation results,
we prefer alternative forms of 
Eqs.~(\ref{eq:twist}) and~(\ref{eq:roll1}),
since the values of these alternative forms do not depend upon the
order of the indices:
\begin{gather} \label{eq:dtheta-roll1}
d\boldsymbol{\theta}^{\text{rel, roll, 1}} \times \mathbf{n} \\
\label{eq:dtheta-twist}
d\theta^{\text{rel, twist}} =
d\boldsymbol{\theta}^{\text{rel}} \cdot \mathbf{n} \;,
\end{gather}
where ``twist'' connotes a relative rotation about
the contact normal $\mathbf{n}$.
\par
\begin{figure}
  \centering
  \includegraphics[scale=0.90]{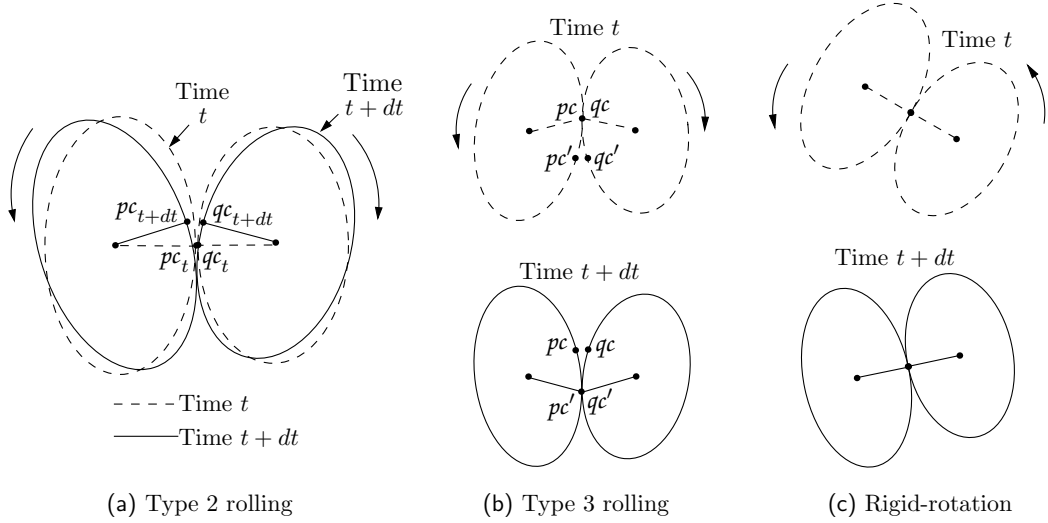}%
  \caption{Particle motions in rolling and in rigid-rotation.}
  \label{fig:rolling}
\end{figure}

A second form of rolling is based upon a particular averaging of
the motions of two material points $\PC$ and $\QC$, 
one attached to each of
the particles at their contact (Fig.~\ref{fig:rolling}a).
The two points are
like two opposing teeth of
inter-meshed gears, with the teeth moving in unison
(in Fig.~\ref{fig:rolling}a, 
the two points lie on the dashed and solid particles
at the times $t$ and $t+dt$, respectively).
The average translation of the two points
(e.g. the opposing gear teeth) is, of course, not objective,
and its magnitude $du^{\mathbf{t}\text{-avr}}$,
in a tangential direction $\mathbf{t}$,
\begin{equation}
du^{\mathbf{t}\text{-avr}} =
\frac{1}{2}
\left[
\left(
d\mathbf{u}^{p} + d\boldsymbol{\theta}^{p}\times \mathbf{r}^{p}
\right) \cdot \mathbf{t}
+
\left(
d\mathbf{u}^{q} + d\boldsymbol{\theta}^{q}\times \mathbf{r}^{q}
\right) \cdot \mathbf{t}
\right]\;,
\end{equation}
is also not objective.
We can produce an objective average 
$du^{\mathbf{t}\text{-roll, 2}}$
by subtracting a common,
rigid-body-like rotation of the pair~\cite{Bagi:2004a}:
\begin{equation} \label{eq:roll2t}
du^{\mathbf{t}\text{-roll, 2}} =
\frac{1}{2}
\left[
(d\boldsymbol{\theta}^{p}\cdot\mathbf{z}^{t})
(\mathbf{r}^{p}\cdot \boldsymbol{\lambda}^{t}) +
(d\boldsymbol{\theta}^{q}\cdot\mathbf{z}^{t})
(\mathbf{r}^{q}\cdot \boldsymbol{\lambda}^{t}) -
\frac{(d\mathbf{u}^{q} - d\mathbf{u}^{p}) \cdot \mathbf{t}}
     {\ell^{\perp t}}
(\mathbf{r}^{p} + \mathbf{r}^{q})\cdot \boldsymbol{\lambda}^{t}
\right] \;,
\end{equation}
where the unit vectors $\mathbf{z}^{t}$ and
$\boldsymbol{\lambda}^{t}$ depend upon the directions of the tangent
vector $\mathbf{t}$ and the branch vector $\mathbf{l}$:
\begin{equation}
\boldsymbol{\lambda}^{t} = \mathbf{l}^{\perp t} / \ell^{\perp t} \;,\quad
\mathbf{z}^{t} = \boldsymbol{\lambda}^{t} \times \mathbf{t}\;,
\end{equation}
with
\begin{equation} \label{eq:lperpt}
\mathbf{l}^{\perp t} = \mathbf{l} - (\mathbf{l}\cdot\mathbf{t})\mathbf{t},
\quad
\ell^{\perp t} = |\mathbf{l}^{\perp t}|\;.
\end{equation}
The quantity $du^{\mathbf{t}\text{-roll, 2}}$
is termed \emph{Type~2 rolling}.
Similar measures of rolling 
are associated with the normal direction $\mathbf{n}$ and
a second tangent direction (say, direction $\mathbf{w}$),
and these measures
can be computed by substituting $\mathbf{n}$ or $\mathbf{w}$
in equations~(\ref{eq:roll2t})--(\ref{eq:lperpt}).
The scalar measures 
of $\mathbf{t}$, $\mathbf{w}$, and $\mathbf{n}$ rolling
are independent of the ordering
of the indices $p$ and $q$.
Unlike Type~1 rolling, which is a rotational quantity, Type~2 rolling
is a translation, which we designate with the symbol $du$
instead of $d\theta$ (\emph{cf}., Eqs.~\ref{eq:roll1} and~\ref{eq:roll2t}).
\par
A third form of rolling is based upon the paths of two contact
points as they travel across the two surfaces while the particles are moving.
For example, when two inter-meshed gears rotate, the two contact points
move from tooth to tooth around the two gears, even as each gear
tooth is moving in the opposite direction.
Figure~\ref{fig:rolling}b shows the pairs of material points
$\PC$--$\QC$ and $\PC'$--$\QC'$,
which are the contact points on the two particles at the times
$t$ and $t'=t+dt$.
\emph{Type~3 rolling} is defined as the average movement
of the two contact points
across the two surfaces.
This form of rolling requires a knowledge of the
local surface curvatures of the two particles
at the contact~\cite{Montana:1988a,Kuhn:2004b} and is given by:
\begin{equation} \label{eq:rolling3}
d\mathbf{u}^{\text{roll, 3}} =
-\left(
\mathbf{K}^{p} + \mathbf{K}^{q}
\right) ^{-1}
\left[
d\boldsymbol{\theta}^{\text{rel}}
\times \mathbf{n}
+
\frac{1}{2}
\left( \mathbf{K}^{p} - \mathbf{K}^{q} \right)
d\overline{\mathbf{u}}^{\text{def}}
\,\right]\;,
\end{equation}
where $\mathbf{K}^{p}$ and $\mathbf{K}^{q}$ are the surface
curvature tensors, and $d\overline{\mathbf{u}}^{\text{def}}$
is the projection of $d\mathbf{u}^{\text{def}}$ 
(defined by Eq.~\ref{eq:udef})
onto the tangent plane of the contact.
Type~3 rolling is clearly objective, since it is a linear
combination of the objective motions $d\mathbf{u}^{\text{def}}$
and $d\boldsymbol{\theta}^{\text{rel}}$.
The vector $d\mathbf{u}^{\text{roll, 3}}$ lies in the
tangent plane of the contact, and its value is independent
of the ordering of indices $p$ and~$q$.
Because it always lies in the tangent plane,
vector $d\mathbf{u}^{\text{roll, 3}}$ is restricted to a 2-dimensional
sub-space of vectors.
The rolling translation $d\mathbf{u}^{\text{roll, 3}}$
should, therefore, be supplemented with an auxiliary
rolling quantity, say $d\theta^{\text{rel, twist}}$,
to complete the full 6-dimensional sub-space of objective motions
(for example, the sub-space formed from the
motions $d\mathbf{u}^{\text{def}}$, $d\mathbf{u}^{\text{roll, 3}}$,
and $d\theta^{\text{rel, twist}}$).
Motions that lie outside of this six-dimensional space, but within the
12-dimensional space of all motions,
are non-objective~\cite{Kuhn:2004c}.
\par
A fourth form of rolling is more abstract than the others.
As has been mentioned,
the set of all objective motions is a 6-dimensional vector sub-space
that lies within the entire 12-dimensional space of motions
for a particle pair.
Rigid-body motions, described more fully in the next 
paragraph, 
also lie in the 12-dimensional space, but they
are not objective.
The fourth type of rolling is defined as the set of objective
motions that are
independent of (and orthogonal to) the vector sub-space
of contact deformation motions in Eq.~(\ref{eq:udef}).
By introducing this fourth form of rolling,
we can project the motions of a particle pair onto three orthogonal
sub-spaces:
the contact deformation sub-space, the Type~4 rolling sub-space, 
and the rigid motion sub-space.
This projection will be applied in 
Section~\ref{sec:def-rot} to explore the three
corresponding sources of deformation in a granular media.
\emph{Type~4 rolling} is defined as the following rotation:
\begin{equation}\label{eq:type4rolling}
\begin{split}
d\boldsymbol{\theta}^{\text{roll, 4}} =
\frac{G}{G^{2}-(\mathbf{s}\cdot\mathbf{l})^{2}}
\left[
(\mathbf{z}\otimes\mathbf{l})(d\mathbf{u}^{q} - d\mathbf{u}^{p})
- 2 \mathbf{s}\times(d\mathbf{u}^{q} - d\mathbf{u}^{p})
\right]
\\
+ \; d\boldsymbol{\theta}^{q} - d\boldsymbol{\theta}^{p}
 - \frac{1}{2}(|\mathbf{r}^{q}|^{2} - |\mathbf{r}^{p}|^{2})\Phi
 - \frac{1}{2}(\mathbf{l}\otimes\mathbf{s}) \Phi \;,
\end{split}
\end{equation}
where
\begin{align}
\Phi &= (1/H^{p})d\boldsymbol{\theta}^{p} +
        (1/H^{q})d\boldsymbol{\theta}^{q} \\
H^{p} &= \mathbf{r}^{p}\cdot\mathbf{l} + 2\\
H^{q} &= -\mathbf{r}^{q}\cdot\mathbf{l} + 2 \\
\mathbf{s} &= \mathbf{r}^{p} + \mathbf{r}^{q}\\
\mathbf{z} &= \mathbf{r}^{p} \times \mathbf{r}^{q} \\
G &= |\mathbf{l}|^{2} + 4\;.
\end{align}
This form of rolling resembles the relative rotation
of Eq.~\ref{eq:relrotation},
in that a permuting of the indices $p$ and $q$ reverses the direction
of $d\boldsymbol{\theta}^{\text{roll, 4}}$.
This characteristic can be corrected by using the following normal
and tangential components:
\begin{align}
d\theta^{\text{roll, 4, n}} &= 
  d\boldsymbol{\theta}^{\text{roll, 4}}\cdot\mathbf{n}\\
\label{eq:type4cross}
d\boldsymbol{\theta}^{\text{roll, 4, t}} &=
  d\boldsymbol{\theta}^{\text{roll, 4}}\times\mathbf{n}\;.
\end{align}
\par
We have also derived a definition of the rigid motion of a
particle pair \cite{Kuhn:2004c}.
The definition gives a type of motion that is purely non-objective.
That is, every motion of two particles (i.e., each element in the
12-dimensional vector space of possible motions)
is the combination of an objective motion and a \emph{purely} non-objective
motion,
which we refer to as the \emph{rigid} component of motion.
The set of all purely non-objective, rigid motions is a
particular sub-space of the 12 degrees of freedom:
the sub-space that is orthogonal
to the sub-space of objective motions.
The \emph{rigid-rotation} is of interest in the current
work (Fig.~\ref{fig:rolling}c), and it is defined as
\begin{equation}\label{eq:rigid}
d\boldsymbol{\theta}^{\text{rigid-rot}} =
\frac{1}{G}
\left[
\mathbf{l} \times \left( d\mathbf{u}^{q} - d\mathbf{u}^{p}\right)
  + 2\left( d\boldsymbol{\theta}^{p} + d\boldsymbol{\theta}^{q} \right)
  + \frac{1}{2} \mathbf{l} \cdot 
       \left( d\boldsymbol{\theta}^{p} + d\boldsymbol{\theta}^{q} \right)
       \mathbf{l}
\right]\;,
\end{equation}
where $G = |\mathbf{l}|^{2} + 4$.
The rigid-translation is defined elsewhere \cite{Kuhn:2004c}.
\subsection{Simulation results: particle rotations and contact deformation}
\label{sec:rot-def}%
Past experimental programs have demonstrated that particle rotations
reduce the stiffness and strength of 2D particle assemblies.
We briefly review this evidence and then provide other experimental
results that characterize the stiffening effect,
with particular attention to 3D assemblies and to the 2D assembly of ovals.
\par
The mechanical effects of particle
rotations have been established in two types of numerical
experiments.  
In one approach, the rotations are numerically restricted with the use of
rotational springs or are prevented 
altogether \shortcite{Bardet:1994a,Calvetti:1997a,Oda:1997a,Iwashita:1998a}.  
The resulting strength and stiffness exceed those in control
experiments with unrestrained rotations.  
In a second approach, measurements
are taken of the incremental effects of the rotations upon the 
contact forces,
which allows a partitioning of the stress increment
into the separate effects of the particle rotations 
and translations \cite{Kuhn:2003h}.  
Such measurements suggest a substantial softening effect of the
particle rotations.
\par
A third type of evidence can be extracted from numerical experiments
by analyzing the contact deformations $d\mathbf{u}^{\text{def}}$
that occur within the assembly (Eq.~\ref{eq:udef}).
These contact deformations produce changes in the contact forces, which
are the source of an assembly's incremental stiffness.
Each contact deformation $d\mathbf{u}^{\text{def}}$ can be separated
into parts that are due to the translations and the rotations of the
pair of particles:
\begin{gather}
d\mathbf{u}^{\text{def}} = 
d\mathbf{u}^{\text{def, trans}} + d\mathbf{u}^{\text{def, rot}},\\
d\mathbf{u}^{\text{def, trans}} = d\mathbf{u}^{q} - d\mathbf{u}^{p}\;,\quad
d\mathbf{u}^{\text{def, rot}} = 
  d\boldsymbol{\theta}^{q} \times \mathbf{r}^{q} -
  d\boldsymbol{\theta}^{p} \times \mathbf{r}^{p}\;.
\end{gather}
We compare the quantities $d\mathbf{u}^{\text{def}}$, 
$d\mathbf{u}^{\text{def, trans}}$, and $d\mathbf{u}^{\text{def, rot}}$
among the thousands of inter-particle contacts 
in each of the five assemblies and at various strains.
The comparisons are coefficients of correlation:
a coefficient of zero indicates no correlation; coefficients of 1 and~$-1$
indicate perfect and perfectly contrary correlations.
The comparison can be broadened by separating $d\mathbf{u}^{\text{def}}$
into its normal and tangential components,
\begin{align}
du^{\text{def, n}}
  &= d\mathbf{u}^{\text{def}}\cdot \mathbf{n} \\
d\mathbf{u}^{\text{def, t}}
  &= d\mathbf{u}^{\text{def}} - 
     du^{\text{def, n}} \mathbf{n} \;.
\end{align}
and computing the corresponding contributions
$du^{\text{def, trans, n}}$, $du^{\text{def, rot, n}}$,
$d\mathbf{u}^{\text{def, trans, t}}$, and $d\mathbf{u}^{\text{def, rot, t}}$.
An analysis of these comparisons leads to the following
conclusions:
\begin{itemize}
\item
Particle rotations have a softening effect by counteracting the
influence of the particle translations.
The translational contributions to contact 
deformation, $d\mathbf{u}^{\text{def, trans}}$, are negatively
correlated with the rotational contributions 
$d\mathbf{u}^{\text{def, rot}}$.
The coefficients of correlation range between $-0.2$ and $-0.65$ across
all particle shapes and strains, and the correlations
are most negative during material softening and at the steady state.
Although one might expect the counter-action of translational 
and rotational contributions to be greater in the
tangential direction than in the normal direction, this is not the case.
For oval and ovoid shapes, the coefficients of correlation between
$d\mathbf{u}^{\text{def, trans, n}}$ and $d\mathbf{u}^{\text{def, rot, n}}$
are even more negative, with values of $-0.7$ to $-0.95$
at large strains.
That is, particle rotations reduce the changes in contact indentation
that would otherwise be produced by particle translations.
\item
At zero strain, the contact deformations $d\mathbf{u}^{\text{def}}$
are closely correlated with the translational contributions
$d\mathbf{u}^{\text{def, trans}}$, with coefficients of
correlation greater than 0.82 for all shapes;
but the contact deformations $d\mathbf{u}^{\text{def}}$ are entirely
uncorrelated with the rotational contributions
$d\mathbf{u}^{\text{def, rot}}$, with coefficients less than 0.01.
Although the rotational fluctuations are uncorrelated with 
contact deformations, the fluctuations are substantial
at zero strain (Section~\ref{sec:part-rot}),
and they reduce material stiffness \cite{Kuhn:2003h}.
The results indicate, however, that a micro-structural approach that
intends to estimate material stiffness
by assuming zero rotation should give a better estimate
of the stiffness at small strains than at large strains.
The correlation between $d\mathbf{u}^{\text{def}}$
and $d\mathbf{u}^{\text{def, trans}}$ is particularly
strong for non-circular and non-spherical shapes
at zero strain,
with correlations greater than 0.95, which indicates
that, at small strain, 
a zero-rotation assumption would be even better suited to 
such shapes.
\item
The opposite situation applies at large strains:  the contact deformations
$d\mathbf{u}^{\text{def}}$ correlate weakly with
the translational contributions $d\mathbf{u}^{\text{def, trans}}$
but correlate strongly with $d\mathbf{u}^{\text{def, rot}}$.
For example, at the steady (critical) state, the correlation 
between $d\mathbf{u}^{\text{def}}$
and $d\mathbf{u}^{\text{def, trans}}$
is only 0.2--0.4.
The correlation between the contact deformations
$d\mathbf{u}^{\text{def}}$ and the rotational contributions
$d\mathbf{u}^{\text{def, rot}}$ is, however, much stronger,
with coefficients between 0.55 and 0.7.
That is, particle rotations assume a dominant role in contact
deformation at large strains.
\end{itemize}
\subsection{Simulation results: contact rolling and rigid-rotation}
Four definitions of rolling were discussed in Section~\ref{sec:definitions}.
The intensities of rolling Types~1, 2 and~3 were measured throughout
the loading experiments on the five assemblies (Table~\ref{table:assemblies}),
and the results are presented in this section.
Type~4 rolling was used to characterize deformations within 
granular sub-regions, as discussed in Section~\ref{sec:def-rot}.
\par
Rolling Types~1, 2, and~3 give different measures of the
rate of rotational interaction at the contacts.  
The Type~1 rolling in Eq.~(\ref{eq:dtheta-roll1}) is a 
relative rotational rate, whereas rolling Types~2 and~3 are translational
rates.
For pairs of circular
and spherical particles of the same size,
the three rates are multiples of each other
(they differ in sign and by multiplicative
factors of~2 and the shared radius).
The three rolling
rates are not equal for non-circular and non-spherical particles.
However, in our tests on oval (2D) and ovoid (3D) shapes, we found
a close correlation among the three types of rolling, with
coefficients of correlation greater than 0.95.
For this reason, we will only report the experimental results
of Type~3 rolling, noting that large differences can
occur among the three rolling measures for individual particle pairs.
\par
Type~3
contact rolling is a vector quantity, a characteristic that must
be considered when averaging these vectors over multiple contacts.
We consider two tangential directions: a $\mathbf{t}$ direction
and a $\mathbf{w}$ direction.
Because the triaxial compressive loading in 3D tests is axisymmetric
about the vertical $\mathbf{e}_{3}$ axis, we use a consistent tangent 
direction $\mathbf{w}$,
\begin{equation}\label{eq:w}
\mathbf{w} = (\mathbf{e}_{3}\times \mathbf{n})
/ |(\mathbf{e}_{3}\times \mathbf{n})\times \mathbf{n}|\;,
\end{equation}
which is horizontal, and a consistent tangent direction $\mathbf{t}$ that
lies in a vertical plane:
\begin{equation}
\mathbf{t} = \mathbf{w} \times \mathbf{n}\;.
\end{equation}
In the rare case of a vertical contact ($\mathbf{e}_{3}\cdot\mathbf{n}=1$),
the vector $\mathbf{w}$ in Eq.~(\ref{eq:w}) does not exist,
and all rolling takes place in the horizontal plane.
In this case, $\mathbf{t}$ and $\mathbf{w}$ are
assigned the $\mathbf{e}_{1}$ and $\mathbf{e}_{2}$ directions.
We also report the rotational twists 
$d\theta^{\text{rel, twist}}$ about the contact normals $\mathbf{n}$
(Eq.~\ref{eq:dtheta-twist}).
The three scalar increments
$d\mathbf{u}^{\text{roll, 3}}\cdot\mathbf{t}$,
$d\mathbf{u}^{\text{roll, 3}}\cdot\mathbf{w}$, and
$d\theta^{\text{rel, twist}}$ have a sign
that is independent of the ordering of the particle indices
($p$ and $q$, Fig.~\ref{fig:particles}).
\par
Tables~\ref{table:rolling-zero} and~\ref{table:rolling-peak} 
show the average intensities
of $\mathbf{t}$- and
$\mathbf{w}$-direction rolling among the thousands of contacts 
in each of the 3D assemblies at
zero strain and at the state of peak stress (at $\epsilon_{33}=-0.02$).
\begin{table}
\centering
\caption{Contact rolling, twist, deformation, and rigid-rotation rates for
3D assemblies at zero strain.}
\begin{tabular}{llddd}
\hline
&&&\multicolumn{2}{c}{Ovoids}\\
\cline{4-5}
Description & Quantity, zero strain & \multicolumn{1}{c}{Spheres} & 
                         \multicolumn{1}{c}{Oblate} & 
                         \multicolumn{1}{c}{Prolate} \\
\hline
$\mathbf{t}$-rolling     &$\mathsf{Std}(d\mathbf{u}^%
                           {\text{roll, 3}}\cdot\mathbf{t})
                          / |d\epsilon_{33}|/\overline{D}$ &
                          0.15 & 0.060 & 0.066 \\
$\mathbf{w}$-rolling     &$\mathsf{Std}(d\mathbf{u}^%
                           {\text{roll, 3}}\cdot\mathbf{w})
                          / |d\epsilon_{33}|/\overline{D}$ &
                          0.11 & 0.043 & 0.051 \\
$\mathbf{n}$-twist       &$\mathsf{Std}(d\theta^%
                          {\text{rel, twist}})
                          / |d\epsilon_{33}| / 4$ &
                          0.13 & 0.048 & 0.045 \\
$\mathbf{t}$-deformation &$\mathsf{Std}(d\mathbf{u}^%
                            {\text{def}}\cdot\mathbf{t})
                          / |d\epsilon_{33}|/\overline{D}$ &
                          0.16 & 0.18 & 0.18 \\
$\mathbf{w}$-deformation &$\mathsf{Std}(d\mathbf{u}^%
                            {\text{def}}\cdot\mathbf{w})
                          / |d\epsilon_{33}|/\overline{D}$ &
                          0.068 & 0.042 & 0.042 \\
Rigid-rotation           &$\mathsf{Std}(|d\boldsymbol{\theta}^%
                            {\text{rigid-rot}}|)
                          / |d\epsilon_{33}|$ &
                         0.26 & 0.12 & 0.11 \\
\hline
\end{tabular}
\label{table:rolling-zero}
\end{table}
\begin{table}
\centering
\caption{Contact rolling, twist, deformation, and rigid-rotation rates for
3D assemblies at the peak state.}
\begin{tabular}{llddd}
\hline
&&&\multicolumn{2}{c}{Ovoids}\\
\cline{4-5}
Description & Quantity, peak state & \multicolumn{1}{c}{Spheres} &
                         \multicolumn{1}{c}{Oblate} &
                         \multicolumn{1}{c}{Prolate} \\
\hline
$\mathbf{t}$-rolling     &$\mathsf{Std}(d\mathbf{u}^%
                           {\text{roll, 3}}\cdot\mathbf{t})
                          / |d\epsilon_{33}|/\overline{D}$ &
                          3.3 & 4.0  & 3.3  \\
$\mathbf{w}$-rolling     &$\mathsf{Std}(d\mathbf{u}^%
                           {\text{roll, 3}}\cdot\mathbf{w})
                          / |d\epsilon_{33}|/\overline{D}$ &
                          3.8 & 4.2 & 3.3  \\
$\mathbf{n}$-twist       &$\mathsf{Std}(d\theta^%
                          {\text{rel, twist}})
                          / |d\epsilon_{33}| / 4$ &
                          3.9 & 3.9  & 2.7  \\
$\mathbf{t}$-deformation &$\mathsf{Std}(d\mathbf{u}^%
                            {\text{def}}\cdot\mathbf{t})
                          / |d\epsilon_{33}|/\overline{D}$ &
                          4.1 & 6.7  & 5.6  \\
$\mathbf{w}$-deformation &$\mathsf{Std}(d\mathbf{u}^%
                            {\text{def}}\cdot\mathbf{w})
                          / |d\epsilon_{33}|/\overline{D}$ &
                          3.6 & 4.8  & 4.0  \\
Rigid-rotation           &$\mathsf{Std}(|d\boldsymbol{\theta}^%
                            {\text{rigid-rot}}|)
                          / |d\epsilon_{33}|$ &
                         8.8 & 10.9 & 8.1 \\
\hline
\end{tabular}
\label{table:rolling-peak}
\end{table}
These rolling intensities are expressed 
as a standard deviation of the rolling rates
in a dimensionless and normalized
form by dividing by the mean particle diameter $\overline{D}$
and by the strain increment $|d\epsilon_{33}|$.
The tables also give the rates of twist
$d\theta^{\text{rel, twist}}$ and the $\mathbf{t}$- and
$\mathbf{w}$-direction rates of contact deformation
(i.e., sliding, Eq.~\ref{eq:udef}).
To compare twist with Type~3 rolling, we divide the 
twist by a factor of~4 to account for the factor of $1/2$ in
Eq.~(\ref{eq:rolling3}) and the ratio 
\mbox{$|\mathbf{r}|/\overline{D}$}, 
which is 
about 1/2 for most particles.
We draw the following conclusions from the results in 
Tables~\ref{table:rolling-zero} and~\ref{table:rolling-peak}:
\begin{itemize}
\item
At small strains,
rolling is smaller among non-spherical particles than among spheres.
At large strain, the rolling rates are about the same for all shapes.
\item
Rolling increases greatly with strain, just as the individual particle
rotations become more rapid (Section~\ref{sec:part-rot}).
Rolling rates at the peak state are from 25 to 100 times greater than
those at the start of loading.
\item
Rolling occurs with a similar intensity in the three directions
$\mathbf{t}$, $\mathbf{w}$, and $\mathbf{n}$.
We view this as an unusual result:  the assemblies were compressed
in the vertical, $\mathbf{e}_{3}$ direction, but the rates
of horizontal and vertical (tangent) rolling are about the same.
\item
Although the rates of contact rolling and contact deformation can 
not be directly compared (rolling is an average movement,
whereas deformation is a movement difference,
Eqs.~\ref{eq:udef} and~\ref{eq:rolling3}),
the results indicate that rolling becomes a relatively more significant
mechanism at larger strains.
\item
As would be expected for axisymmetric vertical compression,
tangential contact deformation (sliding) is greater in the
vertical plane ($\mathbf{t}$-direction)
than in the horizontal, $\mathbf{w}$-direction.
\end{itemize}
\par
We also sought conditions that would promote or inhibit contact rolling,
and we considered the influences of contact orientation,
frictional slipping,
contact force, and local material density.
Our analyses showed that only material density has a significant effect
on the rate of contact rolling:
\begin{itemize}
\item
Figure~\ref{fig:polar} shows the effect of contact orientation on the
rates of contact rolling and contact deformation at the peak stress.
\begin{figure}
  \centering
  \mbox{%
  \subfigure[Orientation $\beta$]%
    {\rule[-5mm]{0mm}{5mm}\includegraphics[scale=0.90]
                                          {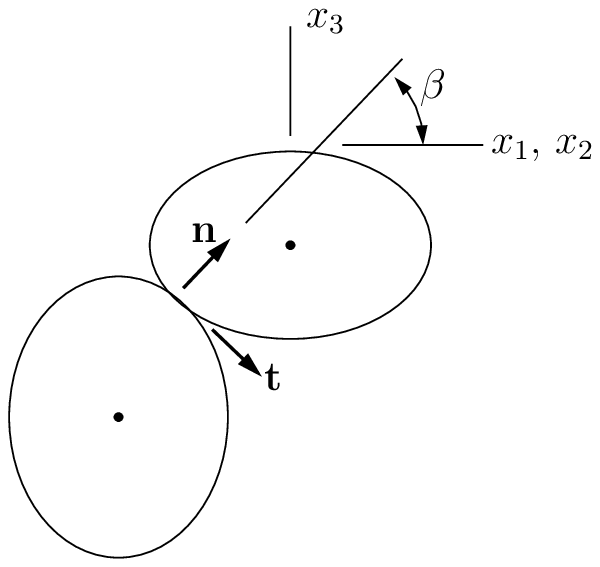}}\quad\quad%
  \subfigure[Rolling magnitude]%
    {\includegraphics[scale=0.90]{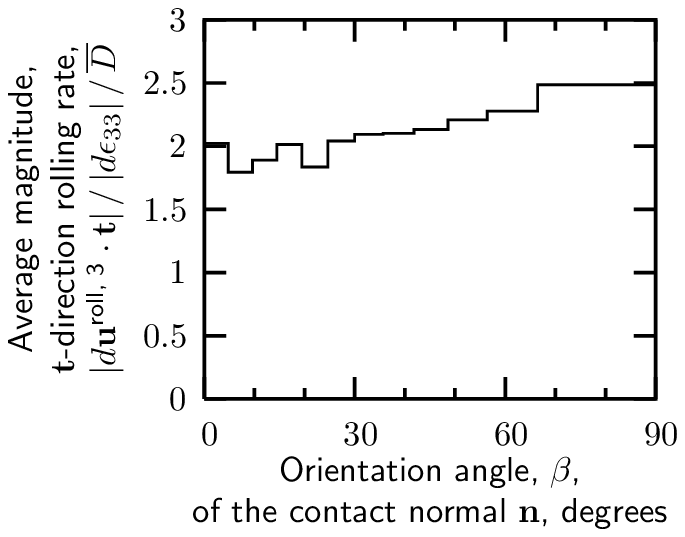}}%
  }\\%
  \mbox{%
  \subfigure[Sliding magnitude]{\includegraphics[scale=0.90]%
    {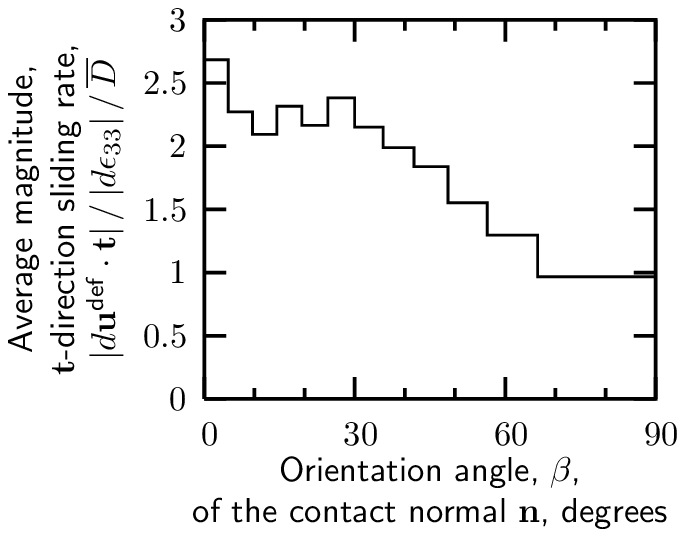}}\quad%
  \subfigure[Sliding]{\includegraphics[scale=0.90]{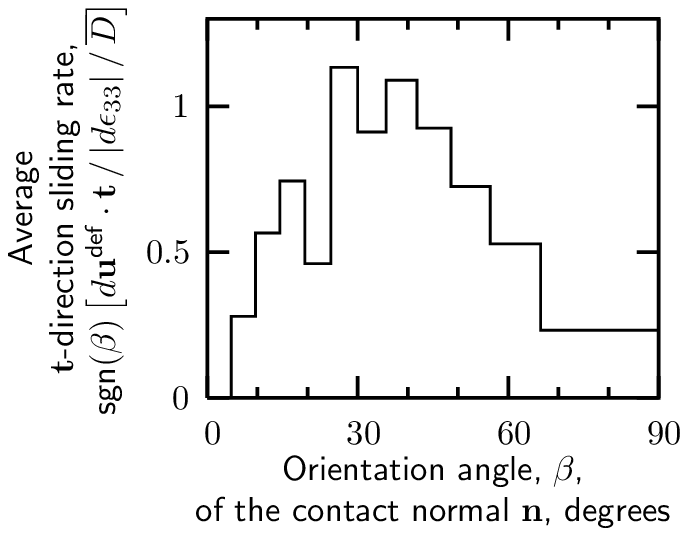}}%
  }
  \caption{Average contact rolling and sliding rates as a function
of the contact orientation $\beta$ at the peak stress.}
  \label{fig:polar}
\end{figure}
For the vertical axisymmetric loading of these simulations,
an orientation angle $\beta$ can be conveniently measured
from the horizontal $\mathbf{e}_{1}$--$\mathbf{e}_{2}$ plane
(Fig.~\ref{fig:polar}a), and symmetries are applied so
that positive and negative $\beta$ angles are folded onto the
range \mbox{0--90$^{\circ}$}.
The average magnitude of the rate of rolling,
$|d\mathbf{u}^{\text{roll, 3}}\cdot\mathbf{t}|$,
is only modestly affected by the orientation $\beta$,
with vertically oriented contacts having slightly greater rolling
rates (Fig.~\ref{fig:polar}b).
This observation can be contrasted with the rates of tangential
contact deformation (i.e., sliding), 
$d\mathbf{u}^{\text{def}}\cdot\mathbf{t}$.
Sliding is, on average, much more vigorous for horizontally
oriented contacts ($\beta=0$), whose average rate magnitude
$|d\mathbf{u}^{\text{def}}\cdot\mathbf{t}|$ is 2.5
times greater than the rate for vertically oriented contacts
(Fig.~\ref{fig:polar}c).
Figure~\ref{fig:polar}d shows the averages of the signed values of
sliding $d\mathbf{u}^{\text{def}}\cdot\mathbf{t}$,
where we have accounted for the anti-symmetric nature of sliding
by plotting the averaged rate
$\text{sgn}(\beta)d\mathbf{u}^{\text{def}}\cdot\mathbf{t}$.
This rate is, of course, zero for horizontal contacts---there is
no preferred sliding direction for a horizontal contact ($\beta=0$).
The maximum average sliding rate occurs at an angle $\beta$
of 30$^{\circ}$--40$^{\circ}$ when the material has been loaded to the
peak state.
\item
Contact rolling is as active among contacts that are undergoing frictional
slipping as among those that are not.
\item
The rate of contact rolling is only slightly affected by the magnitude
of the contact force.  
The coefficient of correlation between the rates
$|d\mathbf{u}^{\text{roll, 3}}|$ and the normal or tangential
contact forces $|f^{\text{n}}|$ or $|\mathbf{f}^{\text{t}}|$
is less than 0.1 at all strains and for all particle shapes.
This observation may be due to the contact law that was used in the 
simulations---linear, elastic
normal and tangential springs---since the incremental
contact rates are influenced by the same incremental contact stiffness, which
is shared by all non-slipping contacts.
\item
The only factor that significantly affects the rate of contact rolling is
the local contact density.
In this regard, we counted the number of contacts
around each pair of particles and measured its effect
on the rolling rate
$|d\mathbf{u}^{\text{roll, 3}}|$ of their common contact.
The coefficient of correlation is between
$-0.05$ and $-0.35$, indicating that the local contact
density inhibits rolling.
\end{itemize}
\par
The rigid-rotation of particle pairs is also an active mode of
interaction (Eq.~\ref{eq:rigid}).
The intensity of this interaction mode is shown in 
Tables~\ref{table:rolling-zero} and~\ref{table:rolling-peak},
which gives the dimensionless standard deviation of
the rigid-rotation magnitudes, 
$|d\boldsymbol{\theta}^{\text{rigid-rot}}|$.
The magnitudes of rigid-rotation are at least as large as
those of contact rolling and deformation, although the larger values
of $|d\boldsymbol{\theta}^{\text{rigid-rot}}|$ in the tables are due,
in part, to our reporting of the full magnitude of the rigid-rotations, 
as compared with the component magnitudes of contact rolling and deformation.
As with contact rolling and contact deformation, 
rigid-rotation is much larger at
the peak state than at zero strain, at which 
rigid-rotation is slightly more active among non-spherical
particles than among spheres.
At larger strains, rigid-rotation has similar intensities for all shapes.
\subsection{Simulation results:  spatial patterning of contact rolling}
\label{sec:gears}
\begin{figure}
  \centering
  \includegraphics[scale=0.600]{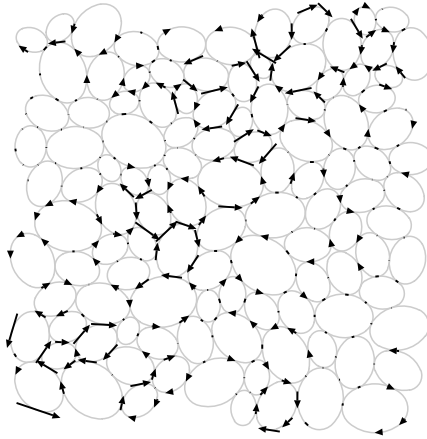}%
  \caption{Type~3 rolling rate vectors 
           $d\mathbf{u}^{\text{roll, 3}}$
           for a small sub-region of ovals during
           biaxial compression.  The figure is taken during material
hardening, when the deviator stress had reached about 70\% of the
peak strength.}
  \label{fig:gears}
\end{figure}

A consistent spatial pattern was observed in the simulated deformation
of 2D assemblies of circular disks and ovals.
This pattern is illustrated in Fig.~\ref{fig:gears},
which shows the rolling vectors among a subset of the 10,816 ovals during
biaxial compression.
The arrows represent the directions of the rolling vectors
$d\mathbf{u}^{\text{roll, 3}}$, and the arrow lengths
are scaled to the magnitudes 
$|d\mathbf{u}^{\text{roll, 3}}|$.
Figure~\ref{fig:gears} reveals that the rolling vectors
around each particle are usually oriented in a common direction, 
either clockwise or counter-clockwise.
For successive pairs of neighboring particles, the
direction of rolling alternates from clockwise to counter-clockwise.
We believe that this pattern is the dominant spatial pattern of
movement in granular materials:
it is observed at all strains and for all particle shapes.
It is present before shear bands form, and it is present within
the shear bands themselves.
This pervasive spatial pattern operates at a length scale of
as small as $1\overline{D}$---a single particle diameter.
In Sections~\ref{sec:curlpattern1} and~\ref{sec:sim-pat-curl}, 
we measure the distances to which this effect extends
from an average particle.
\par
In some respects, this rolling pattern can be likened to the motions among
a set of rotating gears or cogs, since the rolling vectors
around each gear would have a common direction, and this
direction would alternate among neighboring gears.
Although the motions of gears and particles bear some
likeness, the analogy is not entirely appropriate, since
no sliding can occur among intermeshed gears, whereas
rolling between particles occurs simultaneously with contact
deformation (sliding and indentation), as well as with the
rigid-rotations and translations of the particle pairs.
\par
This gear-like pattern 
is more difficult to discern in three-dimensions,
but it seems to be active
in assemblies of spheres and ovoids, although perhaps
to a more muted extent.
As a simple measure of this pattern, 
we considered the subset of 3D particles with five contacts
and found that with 22\% to 33\% of these particles, the rolling
took place in the same direction for all five contacts (either clockwise
or counter-clockwise), when
rolling was measured about the $\mathbf{e}_{1}$ direction.
If the direction of contact rolling was random, the likelihood of
such unanimity would be 6.25\%.
The observation that, on average, 28\% of particles satisfy this
condition suggests that a gear-like rolling pattern 
is active in three-dimensional assemblies.
This notion is also affirmed in
Section~\ref{sec:sim-pat-curl}.
\par
\begin{figure}
  \mbox{%
    \subfigure[Material deformation, \mbox{$\epsilon_{22}=-0.0012$}]
              {\includegraphics[scale=0.88]
                 {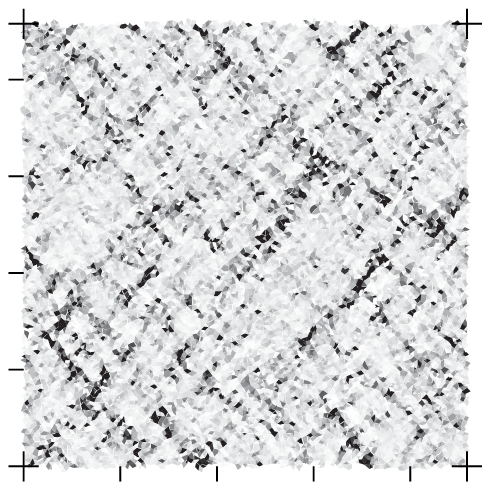}}\quad%
    \parbox[b]{75mm}{\rule{0mm}{10mm}\\%
    \subfigure[Material deformation, \mbox{$\epsilon_{22}=-0.50$}]
              {\includegraphics[scale=0.88]
                 {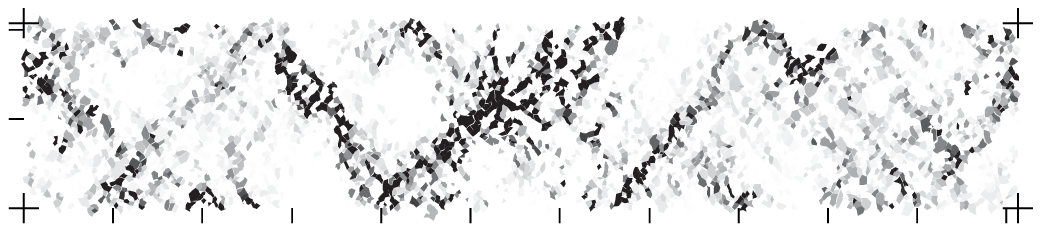}}}%
  }\\
  \mbox{%
    \subfigure[Type~3 rolling, \mbox{$\epsilon_{22}=-0.0012$}]
              {\includegraphics[scale=0.88]
                 {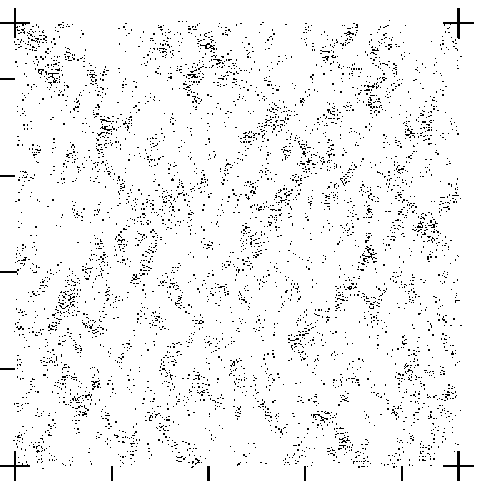}}\quad%
    \parbox[b]{75mm}{\rule{0mm}{10mm}\\%
    \subfigure[Type~3 rolling, \mbox{$\epsilon_{22}=-0.50$}]
              {\includegraphics[scale=0.88]
                 {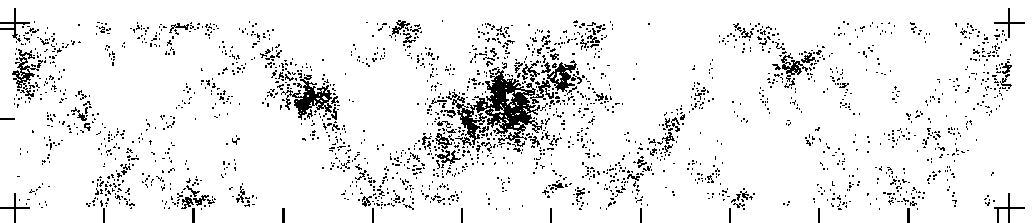}}}%
  }
  \caption{Rates of material deformation and rolling at two strains.
           Figures~(a) and~(c) are during material hardening, when the deviator
           stress had reached about 70\% of the peak stress.
           Figures~(b) and~(d) are at the steady (critical) state.
           The darkest regions in~(a) are deforming more than
           4~times faster than the mean rate (Eq.~\ref{eq:defrate}); 
           the darkest regions
           in~(b) are deforming more than 10~times faster than the mean.}
  \label{fig:def-roll}
\end{figure}

Particle rolling also exhibits spatial patterning at a larger scale.
Figure~\ref{fig:def-roll} shows the spatial
distributions of material deformation and contact rolling
within the oval assembly at two compressive strains:
at the hardening strain of $\epsilon_{22}=-0.0012$ and
at the steady (critical) state of $\epsilon_{22}=-0.50$.
The material deformations in Figs.~\ref{fig:def-roll}a 
and~\ref{fig:def-roll}b
were computed as the deformations within small micro-regions of
material---within the polygonal void cells, which form a complete covering
of a 2D assembly---and the figures show the deformations
in 18,000 and 7,300 such micro-regions
(Section~\ref{sec:def-rot}, 
also see \citeNP{Bagi:1996a} and \citeNP{Kuhn:1999a}).
The shading in Figs.~\ref{fig:def-roll}a and~\ref{fig:def-roll}b
depends upon the contribution of each micro-region's 
rate of deformation tensor
$\mathbf{D}^{\text{micro}}$ to the average assembly
deformation $\mathbf{D}^{\text{average}}$:
\begin{equation}\label{eq:defrate}
\mathbf{D}^{\text{micro}}\cdot\mathbf{D}^{\text{average}} /
|\mathbf{D}^{\text{average}}|^{2}\;,
\end{equation}
where tensor $\mathbf{D}$ is the symmetric part of the
(incremental) velocity gradient.
Micro-bands are visible in the striated texture of Fig.~\ref{fig:def-roll}a,
when the material had been loaded to about 70\% of its peak strength.
Systems of thicker shear bands are seen in Fig.~\ref{fig:def-roll}b,
and these bands are inter-joined across the periodic boundaries.
\par
The rolling vectors $d\mathbf{u}^{\text{roll, 3}}$ 
between oval particles
are represented in
Figs.~\ref{fig:def-roll}c and~\ref{fig:def-roll}d as miniature arrows, 
such as those
in Fig.~\ref{fig:gears}.
Contact rolling is seen to coincide with the localized material deformation,
and rolling is most intense among those particles that lie at
the intersections of crossing shear bands
(Fig.~\ref{fig:gears}d).
\section{A rolling curl}\label{sec:curl}
Because a gear-like pattern of particle motions
appears to dominate in granular materials, 
we propose a measure of the combined extent of rolling translations
at the contacts of a given, central particle.
This measure can be interpreted as a discrete form of material
curl, which may serve as a micro-level state variable of material motion.
In this section, we derive an expression for the average curl
of a rolling vector field and introduce a simpler,
approximate form of this average curl.
Either form can be applied to the rolling of a granular
sub-region (material cells) that is as small as an individual
grain.
Experimental measurements of this \emph{rolling curl}
are presented and discussed.
\subsection{Discrete definition of rolling curl}\label{sec:discrete-curl}
\begin{figure}
  \centering
  \includegraphics[scale=0.90]{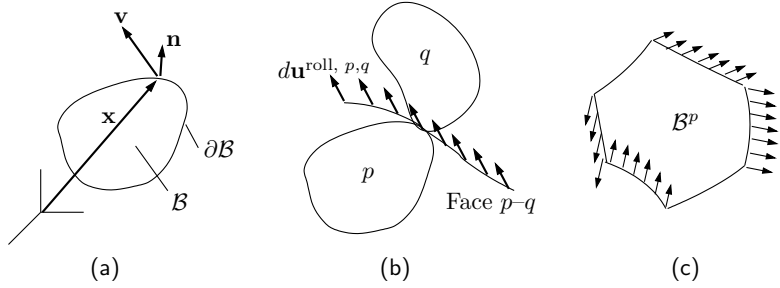}
  \caption{Rolling curl within a material cell.}
  \label{fig:curl}
\end{figure}

The curl of a continuously differentiable vector field $\mathbf{v}$
at a point $\mathbf{x}$ can be defined as the limit of
a surface integral:
\begin{equation}
\boldsymbol{\nabla} \times \mathbf{v}(\mathbf{x}) =
\underset{V\rightarrow 0}{\mathrm{lim}}
\frac{1}{V}
\underset{\partial\mathcal{B}}{\int} \mathbf{n}\times\mathbf{v} \,dS\;,
\end{equation}
where the region $\mathcal{B}$ vanishes around $\mathbf{x}$ and has
a vanishing
boundary surface $\partial\mathcal{B}$ and volume $V$
(Fig.~\ref{fig:curl}a).
In this context, vector $\mathbf{n}$ is the outward unit normal of the
surface $\partial\mathcal{B}$.
The average curl $\overline{\boldsymbol{\nabla} \times \mathbf{v}}$ 
within a finite region $\mathcal{B}$ is
\begin{equation} \label{eq:curl-finite-1}
\overline{\boldsymbol{\nabla} \times \mathbf{v}} =
\frac{1}{V}
\underset{\partial\mathcal{B}}{\int} \mathbf{n}\times\mathbf{v} \,dS\;.
\end{equation}
Because the integral in Eq.~(\ref{eq:curl-finite-1})
applies to the boundary of a finite region, we can relax
the condition of differentiability and require that the field 
$\mathbf{v}$ is continuously differentiable 
on $\mathcal{B}$ except, perhaps, at zero-measure subsets,
such as along surfaces within $\mathcal{B}$ 
and at lines (e.g., polyhedron edges)
or at points (e.g., corners) on the exterior surface $\partial\mathcal{B}$.
For a two-dimensional domain, the volume $V$ would be replaced with
the enclosed area $A$, and integration would be along the region's perimeter.
\par
In order to interpret the curl of rolling in a finite granular region, we
will identify a \emph{material cell} $\mathcal{B}^{\,p}$
as a region (both solid and void) belonging to a single particle
$p$,
such that the union of all material cells forms a non-overlapping
covering (i.e., a partition) of the entire assembly region
(Fig.~\ref{fig:curl}c).
A material cell $\mathcal{B}^{\,p}$
would include all points that are closer to particle $p$
than to any other particle \cite{Bagi:1995a,Bagi:1996a}.
With this scheme,
each material cell is a faceted region, with each face shared
by two neighboring particles.
A rolling vector $d\mathbf{u}^{\text{roll}}$ is associated
with each contact of the particle $p$.
The rolling vector could be based upon either
of rolling Types~1, 2, 3, or~4,
although we later apply a qualifying condition
to the choice of rolling.
Along the faces that correspond to these contacts,
we assign values to a field $d\mathbf{u}^{\text{roll}}(\mathbf{x})$
along each face
(for the purpose of this discrete calculation, 
an assignment of values $d\mathbf{u}^{\text{roll}}(\mathbf{x})$ is
not required within the interior of $\mathcal{B}$,
Eq.~\ref{eq:curl-finite-1}).
In this regard, we can assign a constant rolling vector
field $d\mathbf{u}^{\text{roll}}(\mathbf{x})$
along each contact face:  
the single value of rolling $d\mathbf{u}^{\text{roll}}$
at the face's contact point (Fig.~\ref{fig:curl}c).
For a face that does not correspond to a pair of contacting
particles, the rolling vector field would be zero.
\par
Although we can use this piece-wise constant field 
$d\mathbf{u}^{\text{roll}}(\mathbf{x})$ on the
boundary $\partial\mathcal{B}^{\,p}$, several alternative vector fields 
$d\mathbf{u}^{\text{roll}}(\mathbf{x})$
could also be assigned to points on the boundary.
For example, with rolling Types~2 and~4, a rolling vector 
$d\mathbf{u}^{\text{roll}}$ could be calculated at each point
$\mathbf{x}$ on a face
(Type~3 rolling is, instead, associated with the particles' surface
curvatures at a particular contact point).
A unique definition of the vector field on the interior of $\mathcal{B}$ is
unnecessary: an interpretation of 
definition (\ref{eq:curl-finite-1}) requires only the existence and not the
uniqueness of the field.
\par
After assigning a vector field of 
rolling $d\mathbf{u}^{\text{roll}}(\mathbf{x})$
along the boundary of $\mathcal{B}^{\,p}$,
we can define the corresponding \emph{rolling curl}
as
\begin{equation}\label{eq:roll-curl-discrete}
d\boldsymbol{\rho}^{p} \equiv
\overline{\boldsymbol{\nabla} \times d\mathbf{u}^{\text{roll, }p}} =
\frac{1}{V^{p}}\underset{\partial\mathcal{B}^{\,p}}{\int}
\mathbf{n}\times d\mathbf{u}^{\text{roll, }p} dS\;.
\end{equation}
The assigned boundary vector field $d\mathbf{u}^{\text{roll, }p}(\mathbf{x})$
may change abruptly along the edge lines 
and at the corners between the faces of
material cell $\mathcal{B}^{\,p}$ and its neighboring
cells---the underlying vector field may be discontinuous---but the
existence of the integral~(\ref{eq:roll-curl-discrete})
requires only that the field is continuously differentiable
inside $\mathcal{B}^{p}$ and on the boundary surface $\partial\mathcal{B}^{p}$
except, perhaps, at such sets of zero measure.
\par
The definition of rolling curl in Eq.~(\ref{eq:roll-curl-discrete})
applies to a single material cell $\mathcal{B}^{\,p}$,
but the average curl within a cluster of contiguous cells is
simply the volume-weighted average
\begin{equation}\label{eq:roll-curl-discrete-avg}
d\boldsymbol{\rho}^{\text{cluster}} =
\frac{1}{\sum V^{p}} \sum V^{p} d\boldsymbol{\rho}^{p}
\end{equation}
of the rolling curls in the combined cells.
This intrinsic character of the rolling curl is valid only
if the rolling vector $d\mathbf{u}^{\text{roll}}$ along a 
face is independent of the ordering of the indices, say $p$ and $q$,
of the two particles on either side of the face.
This condition is satisfied by rolling Types~2 and~3,
by the Type~1 rolling in Eq.~(\ref{eq:dtheta-roll1})
and by the cross product
of Type~4 rolling in Eq.~(\ref{eq:type4cross}).
\subsection{Numerical estimation of the rolling curl}
An exact calculation of Eq.~(\ref{eq:roll-curl-discrete})
would require a numerical description of the material 
cells and their boundaries, which are difficult calculations, 
particularly for non-spherical particles.
We have instead used a simpler numerical estimation of the rolling
curl which avoids this difficulty, yet gives a meaningful measure
of the average rate at which neighboring particles roll around
a given, central particle.
The tangent rolling motion $d\mathbf{u}^{\text{roll, }p,q}$ 
between the central particle $p$ and a contacting neighbor $q$
(see Fig.~\ref{fig:curl}b)
can be thought to produce the rotation
\begin{equation}\label{eq:dphipq}
d\pmb{\boldsymbol{\psi}}^{\text{roll, }p,q} =
\frac{d\mathbf{u}^{p,q,\text{ roll}} \cdot \mathbf{y}^{p,q}}
      {\left( \mathbf{r}^{p,q} \times \mathbf{y}^{p,q}
       \right)\cdot \mathbf{w}^{p,q}} \,\mathbf{w}^{p,q}\;,
\end{equation}
where $\mathbf{r}^{p,q}$ is the radial vector from particle $p$ to its
contact with $q$ (the vector $\mathbf{r}^{p}$ in Fig.~\ref{fig:particles}), 
$\mathbf{y}^{p,q}$ is the unit vector in the direction 
of $d\mathbf{u}^{\text{roll, }p,q}$, and $\mathbf{w}^{p,q}$
is the unit vector for 
which the triad $(\mathbf{r}^{p,q},\mathbf{y}^{p,q},\mathbf{w}^{p,q})$
forms a right-hand orthogonal system:
\begin{equation}
\mathbf{y}^{p,q} = \frac{d\mathbf{u}^{\text{roll, }p,q}}
                        {|d\mathbf{u}^{\text{roll, }p,q}|}\;,\quad
\mathbf{w}^{p,q} = \frac{\mathbf{r}^{p,q} \times \mathbf{y}^{p,q}}
                        {|\mathbf{r}^{p,q} \times \mathbf{y}^{p,q}|}\;.
\end{equation}
The average of the imagined rotations~(\ref{eq:dphipq})
at the $M^{p}$ contacts around 
$p$,
\begin{equation}
\overline{d\pmb{\boldsymbol{\psi}}^{p}} =
\frac{1}{M^{p}}\underset{(p,q)}{\sum} 
  d\pmb{\boldsymbol{\psi}}^{p,q,\text{ roll}}\;,
\end{equation}
is the \emph{estimated rolling curl} of the particle.
The estimate can always be applied to convex particles, for which the
tangent rolling vector $d\mathbf{u}^{\text{roll}}$
can not be aligned with the radial vector 
($\mathbf{r}^{p,q}\times\mathbf{y}^{p,q}\neq 0$),
and Eq.~(\ref{eq:dphipq}) will always exist.
\subsection{Simulation results: rolling curl}\label{sec:simcurl}
The objective rolling curl is strongly associated with
the (non-objective) particle rotation:
a rapidly rotating particle will likely be rolling vigorously
at the contacts with its neighbors.
Our results show that the coefficient of correlation 
between a particle's rolling
curl $\overline{d\pmb{\boldsymbol{\psi}}^{p}}$ and its
rotation $\boldsymbol{\theta}^{p}$
is between 0.86 and 0.96 in the biaxial and triaxial loading of 2D and 3D
assemblies.
\par
We found that, on average, 
the rolling curl $\overline{d\pmb{\boldsymbol{\psi}}^{p}}$
is roughly the same for all particle shapes,
but that it is strongly inhibited by contact density.
A three-dimensional particle with 4 contacts will have a
rolling curl that is, on average, two to three times faster than that of
a particle with 8 contacts.
\subsection{Patterning of rolling curl}\label{sec:curlpattern1}
A gear-like spatial pattern of contact rolling was revealed in the
plots of Section~\ref{sec:gears} (Fig.~\ref{fig:gears}).
This pattern can be quantified by considering the rolling
curls $d\pmb{\boldsymbol{\psi}}^{p}$ and $d\pmb{\boldsymbol{\psi}}^{s}$
of pairs of particles (for example, particles $p$ and $s$)
that are separated by a distance.
The intent is to correlate the rolling curl of particle
pairs as a function of the separation between particles.
For example, we would expect that the inner product
$d\pmb{\boldsymbol{\psi}}^{p}\cdot d\pmb{\boldsymbol{\psi}}^{s}$ is, 
on average,
negative for two adjoining particles, since a gear-like pattern
would usually produce opposite curls for two contacting particles.
We will determine if a correlation extends to greater distances.
Rather than measure the geometric distance between
two particles, we instead use a discrete, topologic distance
\cite{Kuhn:2003d}:
the distance $d(p,s)$ between particles $p$ and $s$ 
is the minimum number of branch vectors
(contacts) that must be traveled to reach one particle from the other.
The integer distance $d(p,s)$ would equal~1 for two touching
particles.
Using this metric,
a distance-dependent coefficient of correlation,
$\Psi(\hat{d})$, can be calculated in both 2D and 3D simulations:
\begin{equation}\label{eq:Psi}
\Psi(\hat{d}) = 
\left.
\underset{\{ (p,s):\; d(p,s)= \hat{d}\,\} }{\mathsf{cov}}
\left( d\pmb{\boldsymbol{\psi}}^{p}, d\pmb{\boldsymbol{\psi}}^{s}\right)
\right/
\underset{\{ (p,s):\; d(p,s)= \hat{d}\,\} }{\mathsf{cov}}
\left( d\pmb{\boldsymbol{\psi}}^{p}, d\pmb{\boldsymbol{\psi}}^{p}\right)\;.
\end{equation}
The dimensionless correlation $\Psi(\hat{d})$ is a quotient
of covariances, with each covariance computed from
the entire set of particle pairs $(p,s)$ that are separated
by a distance $\hat{d}$.
The covariance between a set of $N$ vector pairs $(\mathbf{a},\mathbf{b})$
is defined as the average inner product,
\begin{equation}
\mathsf{cov}(\mathbf{a},\mathbf{b}) = \frac{1}{N}
\sum
(\mathbf{a} - \overline{\mathbf{a}})\cdot(\mathbf{b}-\overline{\mathbf{b}})\;,
\end{equation}
where the mean values $\overline{\mathbf{a}}$ and $\overline{\mathbf{b}}$
are subtracted from each instance of $\mathbf{a}$ and $\mathbf{b}$.
The correlation~(\ref{eq:Psi}) at distance $\hat{d}=0$ is simply 1,
since a particle's curl is perfectly correlated with itself, 
for $d(p,p)=0$.
\subsection{Simulation results: patterning of rolling curl}%
\label{sec:sim-pat-curl}
Results are presented in 
Tables~\ref{table:dist-zero} and~\ref{table:dist-peak}
at zero strain and at the peak stress.
The results show an alternating pattern of positive--negative--positive--etc.
correlations, which correspond to a gear-like pattern of rolling.
In the 2D assemblies, this pattern extends to distances
$\hat{d}$ of 6 or more for the average particle.
That is, a particle's motion not only affects its immediate
neighbors, but it also affects a much
larger neighborhood, and the combined effects of all particles
produce a coordinated, patterned system of alternating curls.
The pattern extends to a greater average distance in 2D assemblies
assemblies than in 3D assemblies,
and the affected radius (i.e., the number of
positive--negative alternations) is smaller at zero strain than at 
larger strains.
%
%
%
\begin{table}
\centering
\caption{Zero strain: correlation $\Psi(\hat{d})$ of 
         particle curls at increasing distances,
         $\hat{d}$ (Eq.~\ref{eq:Psi}).}
\begin{tabular}{cddcdd}
\hline
Distance&\multicolumn{2}{c}{2D assemblies}&&\multicolumn{2}{c}{3D assemblies}\\
\cline{2-3}\cline{5-6}
$\hat{d}$ & \multicolumn{1}{c}{Circles} &
            \multicolumn{1}{c}{Ovals} &&
            \multicolumn{1}{c}{Spheres} &
            \multicolumn{1}{c}{Oblate}\\
\hline
0 &  1.0    &  1.0    &&  1.0    &  1.0     \\
1 & -0.55   & -0.29   &&  -0.37  &  -0.21   \\
2 &  0.18   &  0.018  &&  0.080  &   0.017  \\
3 & -0.036  &  0.0058 && -0.0072 &   0.     \\
4 &  0.0059 &  0.0015 &&  0.0008 &   0.0012 \\
5 &  0.0029 &  0.0022 &&  0.     &   0.     \\
6 &  0.     &  0.0014 &&  0.     &   0.     \\
7 &  0.0017 &  0.0010 &&  0.     &   0.     \\
8 &  0.008  &  0.     &&  0.     &   0.     \\
\hline
\end{tabular}
\label{table:dist-zero}
\end{table}
\begin{table}
\centering
\caption{Peak stress: correlation $\Psi(\hat{d})$ of 
         particle curls at increasing distances,
         $\hat{d}$ (Eq.~\ref{eq:Psi}).}
\begin{tabular}{cddcdd}
\hline
Distance&\multicolumn{2}{c}{2D assemblies}&&\multicolumn{2}{c}{3D assemblies}\\
\cline{2-3}\cline{5-6}
$\hat{d}$ & \multicolumn{1}{c}{Circles} &
            \multicolumn{1}{c}{Ovals} &&
            \multicolumn{1}{c}{Spheres} &
            \multicolumn{1}{c}{Oblate}\\
\hline
0 &  1.0    &  1.0    &&  1.0    &  1.0     \\
1 & -0.63   & -0.51   && -0.42   &  -0.33   \\
2 &  0.33   &  0.20   &&  0.14   &   0.091  \\
3 & -0.13   & -0.060  && -0.026  &  -0.013  \\
4 &  0.047  &  0.020  &&  0.0024 &   0.0029 \\
5 & -0.010  & -0.0031 &&  0.0011 &   0.     \\
6 &  0.0051 &  0.0035 &&  0.     &   0.     \\
7 &  0.     &  0.     &&  0.     &   0.     \\
8 &  0.0025 &  0.0010 &&  0.     &   0.     \\
\hline
\end{tabular}
\label{table:dist-peak}
\end{table}
\par
\begin{figure}
\centering
\mbox{%
\subfigure[Circles, $\epsilon_{22}=0$]%
  {\includegraphics[scale=0.90]{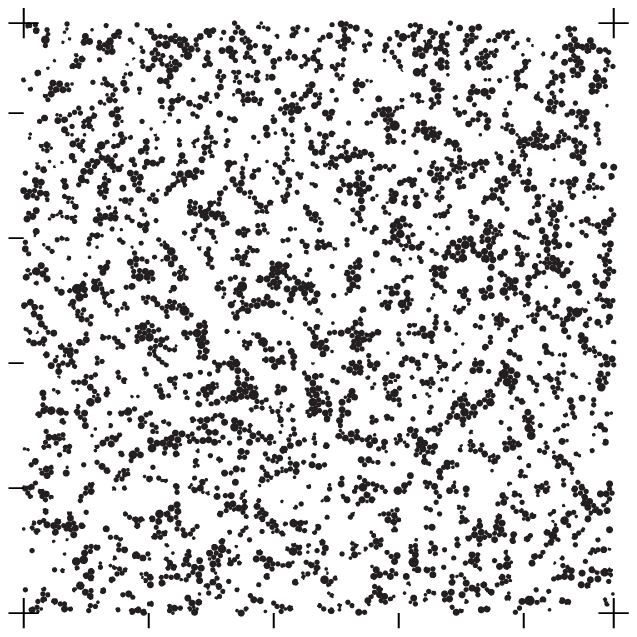}}%
\quad%
\subfigure[Ovals, $\epsilon_{22}=\mbox{}-0.0012$]
  {\includegraphics[scale=0.90]{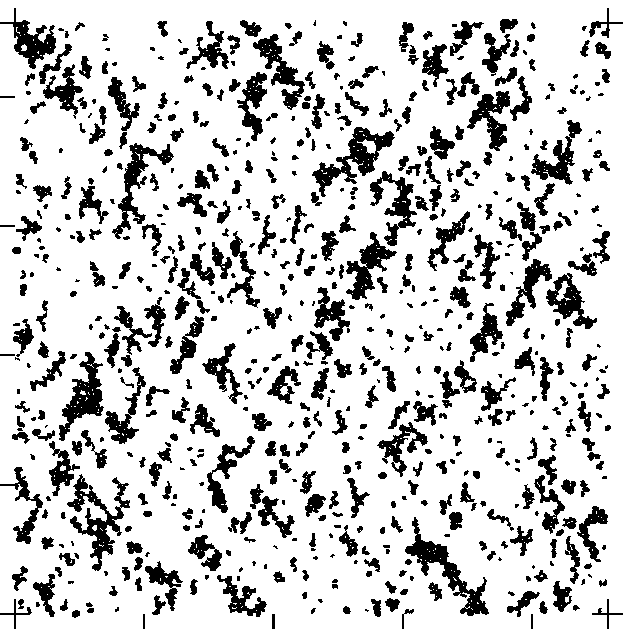}}%
}
\caption{Rolling curl rates of circles and ovals at two strains during
biaxial compression.
The figures depict the curl magnitude
$|\overline{d\pmb{\boldsymbol{\psi}}^{p}}| / |d\epsilon_{22}|$.
In (a), the shaded circles have rates that exceed 0.3;
in (b), the shaded ovals have rates that exceed 0.5.
}
\label{fig:curl-plot}
\end{figure}

The spatial patterning of rolling curl is shown in Fig.~\ref{fig:curl-plot}
for assemblies of circles and ovals.
The monochrome plot gives the magnitudes of rolling curl
$|\overline{d\pmb{\boldsymbol{\psi}}^{p}}|$ for those particles whose
curl exceeds a threshold value.
Figure~\ref{fig:curl-plot}a is for circles at the start of
biaxial compression ($\epsilon_{22}=0$);
Fig.~\ref{fig:curl-plot}b depicts ovals at the hardening
strain $\epsilon_{22}=-0.0012$, when the deviator
stress had reached about 70\% of its peak value.
Both figures show that rolling curl is not uniformly distributed
within an assembly: particles with the most rapid rolling curls
are clustered in small groups of between 10 and~100 particles.
Although these clusters are arranged randomly at small strains, without
any obvious pattern,
at larger strains, the clusters are elongated oblique to
the directions of principal strain.  
Within the clusters, the curl
occurs in the alternating 
positive--negative--positive pattern that was demonstrated
in Tables~\ref{table:dist-zero} and~\ref{table:dist-peak}.
\section{Particle rotations and material deformation}\label{sec:def-rot}
\begin{figure}
  \centering
  \includegraphics[scale=0.90]{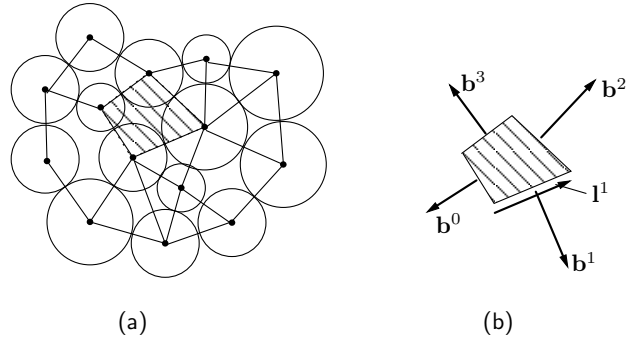}
  \caption{(a) Polygonal void cells in a 2D granular media; (b) rotated
           branch vectors $\mathbf{b}$.}
  \label{fig:graph}
\end{figure}

In this section, we consider the material deformations and rotations
within a granular material and their relation to the particle rotations.
Just as contact deformation is distinct from the deformation of a granular
region, particle rotations can differ from the rotations of
an encompassing material region.
As an example, the rotations of individual particles
were seen to differ greatly from the zero material rotations
of entire assemblies that were undergoing biaxial
or triaxial compression (Section~\ref{sec:part-rot}).
The material deformations and rotations of a granular
sub-region can be computed by tracking the translations
of particles along the boundary of the sub-region \cite{Bagi:1996a}.
In two-dimensional assemblies, the average material
deformation and rotation can be found within small, elemental polygon
micro-regions called \emph{void cells} (Fig.~\ref{fig:graph}a), 
whose $m$ corners lie
at the centers of the adjoining 
particles \cite{Bagi:1996a,Kruyt:1996a,Kuhn:1999a}.
The void cell partition of a 2D granular region 
(such as that in Fig.~\ref{fig:graph}a)
is the dual of the material cell partition that was discussed in
Section~\ref{sec:discrete-curl}
\cite{Satake:1993b}.
The average incremental displacement gradient within
a polygonal region, 
$\overline{\partial\mathbf{u}^{\text{cell}}/\partial\mathbf{x}}$,
is computed as a linear combination of the set of relative
translations $[d\mathbf{u}^{\text{rel}}]$ of the $m$ particle centers:
\begin{equation}\label{eq:P}
\overline{du_{i,j}^{\text{cell}}} = 
\frac{1}{A}
\left[ du_{i}^{\text{rel}} \right]^{\mathrm{T}}
\left[ \mathrm{Q} \right]^{m}
\left[ b_{j} \right]
+
\frac{1}{A}
\left[ q_{i} \right]^{\mathrm{T}}
\left[ b_{j} \right]\;.
\end{equation}
In this equation, $A$ is the area of the $m$-polygon;
$[du_{i}^{\text{rel}}]$ is an $m$-vector
that contains the $\mathbf{e}_{i}$ components of the relative
translations of adjacent corners (particle centers) taken
as pairs around the polygon perimeter;
$[b_{j}]$ is an 
$m$-vector that contains the $\mathbf{e}_{j}$ components of
the $m$ rotated sides of the polygon
(i.e., the rotated branch vectors,
$\mathbf{b} = \mathbf{R}\mathbf{l}$, 
\begin{math}
\mathbf{R}=\left[\begin{smallmatrix}0&1\\-1&0\end{smallmatrix}\right]
\end{math}, as in Fig.~\ref{fig:graph}b);
and $[\mathrm{Q}]^{m}$ is an $m\times m$ coefficient matrix
\cite{Kuhn:1999a}.
For the quadrilateral in Fig.~\ref{fig:graph}b,
the vectors in Eq.~(\ref{eq:P}) are $4\times 1$
and matrix $[ \mathrm{Q}]^{4}$ is $4\times 4$.
The first term on the right of Eq.~(\ref{eq:P})
gives the incremental displacement gradient
$\overline{du_{i,j}^{\text{cell}}}$ when each side (edge, branch vector)
of the polygon deforms in an affine manner.
In granular materials, the deformations along branch vectors 
are concentrated at the 
contacts---a departure from the affine condition.
The second term on the right of Eq.~(\ref{eq:P})
corrects for non-affine deformations along the polygon edges,
with an $m$-vector $[q_{i}]$ that depends upon the translations and
rotations of the boundary particles \cite{Kuhn:2004a}.
\subsection{Simulation results: material rotation}\label{sec:matl-rot}
\begin{figure}
  \centering
  \mbox{%
    \subfigure[At zero strain]%
      {\includegraphics[scale=0.90]{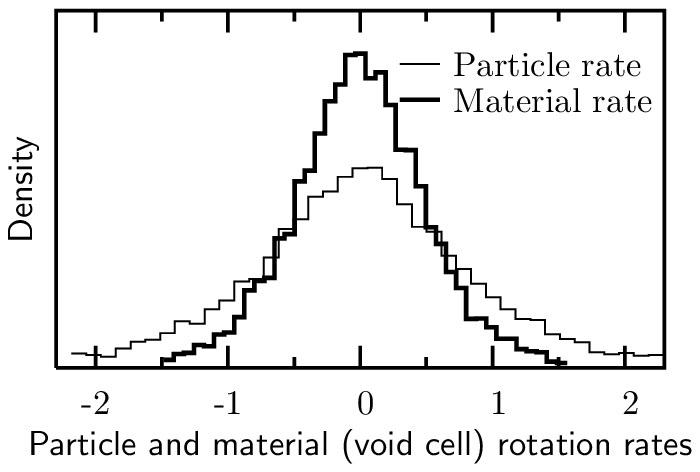}}\ %
    \subfigure[At peak stress]%
      {\includegraphics[scale=0.90]{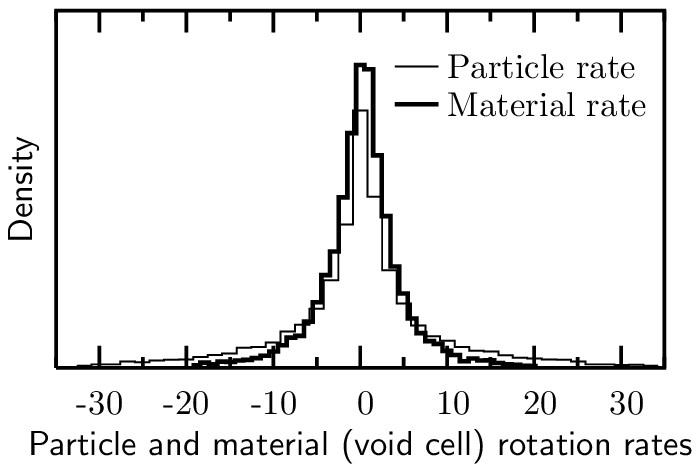}}%
  }
  \caption{Particle rotations and material (void cell) rotations in the
           2D disk assembly.  Probability densities are for the
           dimensionless particle rotations, 
           $d\theta^{p}_{3} / |d\epsilon_{22}|$, and for the
           dimensionless void cell rotations,
           $(\overline{du_{2,1}^{\text{cell}}} -
             \overline{du_{1,2}^{\text{cell}}}) / |d\epsilon_{22}|$.}
  \label{fig:cellRotate}
\end{figure}

The histograms of Fig.~\ref{fig:cellRotate} compare the
particle rotations and void cell (material) rotations
of the assembly of circular disks at two strains.
Although the assembly rotation is zero under biaxial
loading conditions,
material rotation is non-uniform when viewed at a micro-scale,
even at the start of loading.
At zero strain, about 7\% of the void cells are rotating at rates
$du_{2,1} - du_{1,2}$ that exceed the rate of vertical compression,
$\epsilon_{22}$.
These material rotations are largely organized in a micro-band
patterning \cite{Kuhn:1999a}.
At the peak stress, material rotations are quite large, with about 7\% of
the void cells rotating at rates greater than 10 times the 
rate of vertical compression.
At small strain, the material (void cell) rotations are, on average, smaller
and less varied than the particle rotations:
the distributions of material rotation have greater peaks and
smaller tails than those of the particle rotation rates.
At the peak state, the distributions of material rotation and particle
rotation are similar, with particle rotations being only modestly larger
in their magnitude and scatter.
\subsection{Partitions of material deformation}\label{sec:void-cell-partitions}
Several investigators have measured the deformations within such
polygonal micro-regions \cite{Kruyt:1996a,Kuhn:1999a,Lanier:2000a,Roux:2003a}.
In the current study, we use Eq.~(\ref{eq:P}) to 
investigate two partitions of the void cell deformations, to gain
a better understanding of the source and nature of granular deformation.
These partitions were applied in the simulations of biaxial compression
of the 2D assembly of circular disks.
This approach is a counterpart to the use of stress partitions
to explore force transmission in granular materials
\cite{Cundall:1983a,Kuhn:2003h}.
The deformation partitions are as follows.
\begin{enumerate}
\item
In the first partition, we separate the effects of relative 
disk translations that are aligned with and that are 
perpendicular to the contact
normals.
Each of the $m$ relative movements $d\mathbf{u}^{\text{rel}}$
in Eq.~(\ref{eq:P}) is the relative translation
of two particle centers, say $p$ and $q$, at adjacent corners of a 
polygon:
\begin{equation}\label{eq:R}
du_{1}^{\text{rel, }pq} = du^{q}_{1} - du^{p}_{1}\text{\quad and\quad}
du_{2}^{\text{rel, }pq} = du^{q}_{2} - du^{p}_{2} \;.
\end{equation}
A relative translation $d\mathbf{u}^{\text{rel, }pq}$
can be separated into components that are parallel and perpendicular to
the contact normal $\mathbf{n}$:
\begin{equation}\label{eq:void-cell-move-1}
d\mathbf{u}^{\text{rel, }pq} =
du^{\text{rel, }pq\text{, n}}\mathbf{n} +
du^{\text{rel, }pq\text{, t}}\mathbf{t}\;.
\end{equation}
The separate effects of these two parts are investigated by
substituting them individually in Eq.~(\ref{eq:P}) and measuring
the corresponding deformations during biaxial compression:
\begin{equation}\label{eq:defnt}
\overline{du_{i,j}^{\text{cell}}} =
\overline{du_{i,j}^{\text{cell, n}}} +
\overline{du_{i,j}^{\text{cell, t}}}\;,
\end{equation}
which yields
an additive partition of the void cell deformations.
\item
In the second partition, we separate the effects of the contact deformation,
the Type~4 contact rolling, and the rigid-rotation of each pair
of particles
around the polygon's perimeter.
These three types of motions were noted as forming orthogonal
sub-spaces within the complete set of possible motions for a particle
pair (Section~\ref{sec:definitions}).
In two dimensions, the contact deformation, contact rolling, and rigid motion
can be computed from the particle movements
$d\mathbf{u}^{p}$, $d\mathbf{u}^{q}$,
$d\boldsymbol{\theta}^{p}$, and $d\boldsymbol{\theta}^{q}$, as the matrix
product
\begin{equation}\label{eq:Amatrix}
\begin{bmatrix}
du_{1}^{\text{def, }pq} \\
du_{2}^{\text{def, }pq} \\
d\theta^{\text{roll, 4, }pq} \\
d\theta_{3}^{\text{rigid-rot, }pq} \\
du_{1}^{\text{rigid-trans, }pq} \\
du_{2}^{\text{rigid-trans, }pq}
\end{bmatrix}
=
[\mathrm{A}]_{6\times 6}
\begin{bmatrix}
du_{1}^{p} \\
du_{2}^{p} \\
d\theta_{3}^{p} \\
du_{1}^{q} \\
du_{2}^{q} \\
d\theta_{3}^{q}
\end{bmatrix}\;,
\end{equation}
where the $6\times 6$ matrix $[\mathrm{A}]$ effects a linear combination
of the six particle motions that are collected in the $6\times 1$ vector
on the right \cite{Kuhn:2004c}.
The contents of the first four rows of $[\mathrm{A}]$
are manifestations of Eqs.~(\ref{eq:udef}), 
(\ref{eq:type4rolling}), and~(\ref{eq:rigid}).
\par
The relative motions $d\mathbf{u}^{\text{rel}}$
in Eq.~(\ref{eq:P}) are the differences in the translations
of pairs of adjoining particles, say $p$ and $q$, along the
polygon sides, as in Eq.~(\ref{eq:R}).
By inverting matrix $[\mathrm{A}]$ and multiplying by
the left side of Eq.~(\ref{eq:Amatrix}),
we can express the relative translation~(\ref{eq:R})
as a selective combination of contact deformation,
contact rolling, and rigid-rotation effects:
\begin{align}
\label{eq:du1relpq}
du_{1}^{\text{rel, }pq} &=
\left(
\left[ \mathrm{A}^{-1}_{4,1..4}\right] -
\left[ \mathrm{A}^{-1}_{1,1..4}\right]
\right)
\begin{bmatrix}
du_{1}^{\text{def, }pq} \\
du_{2}^{\text{def, }pq} \\
d\theta^{\text{roll, 4, }pq} \\
d\theta_{3}^{\text{rigid-rot, }pq}
\end{bmatrix} \\
\label{eq:du2relpq}
du_{2}^{\text{rel, }pq} &=
\left(
\left[ \mathrm{A}^{-1}_{5,1..4}\right] -
\left[ \mathrm{A}^{-1}_{2,1..4}\right]
\right)
\begin{bmatrix}
du_{1}^{\text{def, }pq} \\
du_{2}^{\text{def, }pq} \\
d\theta^{\text{roll, 4, }pq} \\
d\theta_{3}^{\text{rigid-rot, }pq}
\end{bmatrix}\;,
\end{align}
where we have used
the rows 1, 2, 4, and~5 and the columns 1, 2, 3, and~4
of the inverse matrix $[\mathrm{A}]^{-1}$ to
find the two components of relative translation,
$du_{1}^{\text{rel, }pq}$ and $du_{2}^{\text{rel, }pq}$.
For example, the single effect of contact rolling upon the
relative translation $du_{1}^{\text{rel, }pq}$ is
\begin{equation}\label{eq:du1relroll}
du_{1}^{\text{rel, roll, }pq} =
\left(
A^{-1}_{4,3} - A^{-1}_{1,3}
\right)
d\theta^{\text{roll, 4, }pq}\;.
\end{equation}
The cumulative rolling effect upon the void cell deformation,
$\overline{du_{1,j}^{\text{cell, roll}}}$,
can then be found by assembling
the corresponding effects (\ref{eq:du1relroll}) of each
particle pair into an $m$-vector on the right of Eq.~(\ref{eq:P}).
\par
This approach yields the separate contributions
of contact deformation, contact rolling, and rigid-rotation to the material
deformation within a void cell:
\begin{equation}\label{eq:dudrr}
\overline{du_{i,j}^{\text{cell}}}
=
\overline{du_{i,j}^{\text{cell, def}}} +
\overline{du_{i,j}^{\text{cell, roll}}} +
\overline{du_{i,j}^{\text{cell, rigid-rot}}}\;.
\end{equation}
The rigid-translations in Eq.~(\ref{eq:P}) make no contribution
to $\overline{du_{i,j}^{\text{cell}}}$, since
their contribution to the relative motion
$d\mathbf{u}^{\text{rel, }pq}$ is zero for each
particle pair
(compare Eq.~\ref{eq:Amatrix} 
with Eqs.~\ref{eq:du1relpq} and~\ref{eq:du2relpq}, 
where the motions
$du_{1}^{\text{rigid-trans, }pq}$ and $du_{1}^{\text{rigid-trans, }pq}$
are removed for this reason).
The additive form in Eq.~(\ref{eq:dudrr})
is warranted by the nature of matrix $[\mathrm{A}]$
in Eq.~(\ref{eq:Amatrix}):
the contact deformation, rolling, and rigid-rotation occur in orthogonal
``directions'' (the product of $[\mathrm{A}]$
and its transpose $[\mathrm{A}]^{\mathrm{T}}$ is block diagonal),
so that the three motions contribute independently to the relative
translations $d\mathbf{u}^{\text{rel, }pq}$ in Eq.~(\ref{eq:R}).
\par
The contribution of contact deformations to the void cell (material)
deformation, $\overline{du_{i,j}^{\text{cell, def}}}$,
can be further separated into the following three parts:
a part $\overline{du_{i,j}^{\text{cell, def, n}}}$
due to the relative particle motions that are aligned with
the contact normal;
a part $\overline{du_{i,j}^{\text{cell, def, t-elast}}}$
due to the tangential sliding motions that produce
elastic deformations at the contacts (i.e., in the contact springs);
and a part $\overline{du_{i,j}^{\text{cell, def, t-slip}}}$
due to frictional slip at the contacts.
The first part,
$\overline{du_{i,j}^{\text{cell, def, n}}}$,
is equal to the part $\overline{du_{i,j}^{\text{cell, n}}}$
of the first partition, Eq.~(\ref{eq:defnt}).
\end{enumerate}
\subsection{Simulation results: material deformation}
The two partitions of Eqs.~(\ref{eq:defnt}) and~(\ref{eq:dudrr})
were investigated in the numerical simulation of 2D disks in biaxial
compression, and the results are
assembled in Tables~\ref{table:def-nt} and~\ref{table:def-drr}.  
The tables give deformation rates at zero strain and at the peak stress.
Both volumetric (dilatancy) and distortional rates are reported.
Our analysis follows.
\begin{table}
\centering
\caption{Cumulative effect of the normal and tangential particle movements 
         on the material deformation.  
         The results are for simulated biaxial
         compression of a 2D assembly of circular disks.  
         The assembly deformation is a area-weighted tally 
         of the micro-deformations within void cell regions.}
\begin{tabular}{clcc}
\hline
&&\multicolumn{2}{c}{State}\\
\cline{3-4}
&& Zero & Peak \\
\hline
&Dilation & & \\
A&\quad Normal movement, 
  $\sum(\overline{du_{1,1}^{\text{cell, n}}} + 
    \overline{du_{2,2}^{\text{cell, n}}})/|d\epsilon_{22}|$ & 
  $-0.81$ & $-0.04$\\
B&\quad Tangential movement,
  $\sum(\overline{du_{1,1}^{\text{cell, t}}} +
    \overline{du_{2,2}^{\text{cell, t}}})/|d\epsilon_{22}|$ & 
  $-0.06$ &  $\;\;\;0.60$\\
C&\quad Total dilation, $(d\epsilon_{11} + d\epsilon_{22})/|d\epsilon_{22}|$%
  \hfill$\sum=$& 
  $-0.87$ & $\;\;\;0.56$ \\
&Distortion & & \\
D&\quad Normal movement,
  $\sum(\overline{du_{1,1}^{\text{cell, n}}} -
    \overline{du_{2,2}^{\text{cell, n}}})/|d\epsilon_{22}|$ &
  $\;\;\;0.53$ & $\;\;\;0.04$\\
E&\quad Tangential movement,
  $\sum(\overline{du_{1,1}^{\text{cell, t}}} -
    \overline{du_{2,2}^{\text{cell, t}}})/|d\epsilon_{22}|$ &
  $\;\;\;0.60$ &  $\;\;\;2.52$\\
F&\quad Total distortion, $(d\epsilon_{11} - d\epsilon_{22})/|d\epsilon_{22}|$%
  \hfill$\sum=$&
  $\;\;\;1.13$ & $\;\;\;2.56$ \\
\hline
\end{tabular}
\label{table:def-nt}
\end{table}
\begin{table}
\centering
\caption{Cumulative effect of the contact deformation, rolling, and
         rigid movements
         on the material deformation.
         The results are for simulated biaxial
         compression of a 2D assembly of circular disks.
         The assembly deformation is a area-weighted tally 
         of the micro-deformations within void cell regions.}
\begin{tabular}{cldd}
\hline
&&\multicolumn{2}{c}{State}\\
\cline{3-4}
&& \multicolumn{1}{c}{Zero}& \multicolumn{1}{c}{Peak}\\
\hline
&Dilation & & \\
&\quad Contact deformation,
  $\sum(\overline{du_{1,1}^{\text{cell, def}}} +
    \overline{du_{2,2}^{\text{cell, def}}})/|d\epsilon_{22}|$ & & \\
A&\quad\quad Normal movements& -0.81 & -0.04 \\
B&\quad\quad Tangential, elastic sliding& -0.03 & 0. \\
C&\quad\quad Tangential, frictional slip& 0. & -0.14 \\
D&\quad Rolling,
  $\sum(\overline{du_{1,1}^{\text{cell, roll}}} +
    \overline{du_{2,2}^{\text{cell, roll}}})/|d\epsilon_{22}|$ &
   0. & -0.05 \\
E&\quad Rigid-rotation,
  $\sum(\overline{du_{1,1}^{\text{cell, rigid-rot}}} +
    \overline{du_{2,2}^{\text{cell, rigid-rot}}})/|d\epsilon_{22}|$ &
  -0.35 & 0.79 \\
F&\quad Total dilation, $(d\epsilon_{11} + d\epsilon_{22})/|d\epsilon_{22}|$%
  \hfill$\sum=$&
  -0.87 & 0.56 \\
&Distortion & & \\
&\quad Contact deformation,
  $\sum(\overline{du_{1,1}^{\text{cell, def}}} -
    \overline{du_{2,2}^{\text{cell, def}}})/|d\epsilon_{22}|$ & & \\
G&\quad\quad Normal movements& 0.53 & 0.04 \\
H&\quad\quad Tangential, elastic sliding& 0.28 & 0. \\
I&\quad\quad Tangential, frictional slip& 0. & 0.72 \\
J&\quad Rolling,
  $\sum(\overline{du_{1,1}^{\text{cell, roll}}} -
    \overline{du_{2,2}^{\text{cell, roll}}})/|d\epsilon_{22}|$ &
  -0.02 & -0.12 \\
K&\quad Rigid-rotation,
  $\sum(\overline{du_{1,1}^{\text{cell, rigid-rot}}} -
    \overline{du_{2,2}^{\text{cell, rigid-rot}}})/|d\epsilon_{22}|$ &
  0.33 & 1.93 \\
L&\quad Total dilation, $(d\epsilon_{11} - d\epsilon_{22})/|d\epsilon_{22}|$%
  \hfill$\sum=$&
  1.13 & 2.56 \\
\hline
\end{tabular}
\label{table:def-drr}
\end{table}
\begin{itemize}
\item
Changes in the contact indentations 
tend to reduce an assembly's volume (Table~\ref{table:def-nt}, row~A).
Dilation, however, is largely the result of particle movements
that are perpendicular to the contact normals (movements
$du^{\text{rel, }pq\text{, t}}$ in Eq.~\ref{eq:void-cell-move-1};
Table~\ref{table:def-nt}, row~B).
At both small and large strains, the normal movements
produce compression, although the compressive influence
is reduced at larger strains.
The effect of the normal movements
is counteracted by the increasingly dilatant influence of the
tangential movements, which produces the net material
dilation at large strains.
\item
At small strains, both the normal and tangent
motions produce material distortion, and they do so in
roughly equal measure 
(Table~\ref{table:def-nt}, rows D and~E).
At large strains, the distortion is almost entirely attributed
to inter-particle movements that are tangent to the contacts.
\item
At zero strain, volume change is almost entirely attributed
to contact deformation, with almost no contribution from
the other forms of relative particle translation
(Table~\ref{table:def-drr}, rows~A--C).
This result suggests that a mean-strain approach to estimating
the bulk modulus would be particularly successful at small strains.
At large strains, volume change results primarily from the
rigid-rotations of particle pairs
(Table~\ref{table:def-drr}, row~E).
The dominant effect of tangent motions on the dilation at large strains was
noted in the first observation.
For circular disks, the tangent motions are expressed
as combinations of contact sliding (i.e., tangent contact deformation,
both elastic and slip), contact rolling, and rigid-rotation.
Tables~\ref{table:def-nt} and~\ref{table:def-drr} suggest that
contact sliding has a small compressive effect, but that
this effect is counteracted by the dilatant effect of the rigid-rotations.
\item
At small strains, distortion mainly results from contact deformations
(both normal and tangential), but also from the rigid-rotations of particle
pairs (Table~\ref{table:def-drr}, rows~G--I and~K).
At large strains, most distortion is attributed to
the rigid-rotations, with a smaller contribution from contact sliding.
This smaller contribution of contact sliding is predominantly attributed
to frictional slip, rather than to the elastic contact deformations
(either normal or tangential).
\item
As expected, frictional slip has very little influence on deformation
at the start of loading.
At large strains, slip has significant compressive and distortional
effects.
\end{itemize}
\par
Table~\ref{table:def-drr} also shows that contact rolling contributes little
to the overall deformation of the assembly.
The small cumulative influence of rolling does not mean, however,
that contact rolling is not active in material deformation.
At the peak state,
the local, individual void cell contributions of rolling,
$\overline{du_{i,j}^{\text{cell, roll}}}$,
were large and varied,
but the local positive
and negative contributions canceled each other in our tally of the
cumulative effect.
\section{Conclusion}
The paper has focused on numerical experiments to analyze 
rotations and rolling in granular assemblies, and on their
role in the overall material deformation.
Different scales were analyzed:
individual particle rotations, contact rolling
between pairs of particles, material rotations of void cells,
and rotational patterns on extended domains.
The most important findings are as follows:
\begin{itemize}
\item
Confirming the results of \citeN{Jenkins:2003a} 
and \shortciteN{Calvetti:1997a},
the average \emph{particle rotation} can differ slightly from
the mean continuum rotation of an assembly, although the
average particle rotation is small in comparison to the rotational fluctuations
of individual particles.
The particle shape does not have an appreciable 
effect on the average of the particle rotations,
but, especially prior to the peak stress state, 
the statistical scatter is larger in the case of circles and spheres than
for ovals and ovoids.
The most rapidly rotating particles are aligned in chain-like patterns.
Particle rotations become more rapid with increasing strain until the
peak stress is attained.
\item
The interaction between two contacting particles 
is a combination of contact deformation, \emph{contact rolling}, and
a \emph{rigid-rotation} of the pair.
Different measures of contact rolling were introduced.
These measures are closely correlated with each other, so that each could
serve as the basis of a kinematical state variable in future constitutive
theory.
Concerning the spatial distribution of contact rolling, we found
that (a)~the directions of the rolling vectors around an
individual particle correlate strongly with each other; and (b)~except
at the initial strain, the most rapidly rolling contacts form strip-like
domains that coincide with the most rapidly deforming regions of the assembly.
Particle rotations have a softening effect, by reducing the contact
deformation that would otherwise be produced by the particle translations.
\item
A micro-level state variable belonging to the individual grains,
called the \emph{rolling curl} was defined.
Using this quantity, we analyzed spatial correlations 
between the rolling motions of particles, and the results give
clear evidence of a gear-like pattern of rolling motions.
\item
\emph{Material rotations} were calculated as the anti-symmetric part
of the velocity gradient within material void cells, and the material rotations
were compared with the particle rotations.
At small strain, the material rotations were smaller and less scattered;
but at the peak stress state, the material rotations were nearly
as large as the particle rotations in both their averages and their
dispersions.
\item
The \emph{deformations of void cells} were partitioned in two different ways,
and the significance of the deformation parts were appraised during
the simulated deformation processes.
Though contact rolling motions made a small contribution to the overall
deformation of the assemblies, the rolling contributions in the individual
voids were large.
\end{itemize}
\par
In future experimental studies we would like to learn how the rolling
and deformation behaviors are influenced by particle characteristics,
such as contact friction, and by geometrical features, such as
grain size distribution and particle shape.
We also plan to analyze unloading and cyclic loading processes, to
supplement the monotonic loadings that were exclusively considered
in the present paper.
We expect that the results of the current and future studies will provide
a sound basis for developing a constitutive theory that 
incorporates the effects of rolling motions and particle rotations.
\section*{Acknowledgement}
The work was funded in part by the grant OTKA~31889.  This support is
gratefully acknowledged.
%
%
%
%

\end{document}